# The information path functional approach for solution of a controllable stochastic problem

**Vladimir S. Lerner**

13603 Marina Pointe Drive, C-608, Marina Del Rey, CA 90292, USA, lernervs@gmail.com

***Abstract***

We study a stochastic control system, described by Ito's controllable equation, and evaluate the solutions by an entropy functional (EF), defined by the equation's functions of controllable drift and diffusion.

Considering a control problem for this functional, we solve the EF control's variation problem (VP), which leads to both a dynamic approximation of the process' entropy functional by an information path functional (IPF) and to an information dynamic model (IDM) of the stochastic process.

The IPF variation equations allow finding the optimal control functions, applied to both stochastic system and the IDM for a joint solution of the identification and optimal control problems, combined with a state's consolidation.

In this optimal dual strategy, the IPF optimum predicts each current control action not only in terms of a total functional path goal, but also by setting for each following control action the renovated values of this functional's controllable drift and diffusion, identified during the optimal movement, which concurrently correct this goal.

The VP information invariants allow optimal encoding of the identified dynamic model's operator and control.

The introduced method of cutting off the process by applying an impulse control estimates the cut off information, accumulated by the process' *inner connections* between its states.

It has shown that such a functional information measure contains more information than the sum of Shannon's entropies counted for all process' separated states, and provides information measure of Feller's kernel.

Examples illustrate the procedure of solving these problems, which has been implemented in practice.

## *1. Introduction*

Optimal control of stochastic systems still remains an actual problem in control theory and practice, which involves identification of the random processes by a corresponding dynamic model and applying optimal control to both the system and its dynamic model.

Solution of this problem for an *information dynamic system* not only generalizes a potential solution, but would also allow specialize the universal nature information model for each particular applications.

The optimal control of a stochastic system [2, 3, 4, 15, 21, 38, other] is accompanied by a selection (filtering) of a suitable (needed) *dynamic* signal by a control, minimizing a random noise.

A practical interest represents the system information model's restoration during a joint solution of both the identification and the optimal control problems, allowing to obtain a *current* optimal model of the observed process, which could have been changed during observation.

Many years have passed since the remarkable publications [1,12,18] but *significant new* results in this area have not been achieved. A *complete solution of this problem is unknown*.

The conventional methods for the model's identification by system's observed data [11, 30, 31, 33, other] do not use the *concurrent* identification, combined with optimal control.

We are seeking this problem solution through a new approach based on the optimization of the *entropy path functional* (EF) defined on trajectories of a stochastic system by analogy to R.P. Feynman's *path functional* (FPF) [13,14], which includes a variation principle to obtain the equations of quantum mechanics.

Following Feynman's ideology, we formulate and solve the EF variation problem (VP) leading to the EF *dynamic* form, as an *information path functional* (IPF), defined on the *dynamic* trajectories of a sought information dynamic model (IDM) for the stochastic system.

The approach is applied to a controllable random system described by the *structure of* a multi-dimensional stochastic differential equation, whose functions of a controllable drift and diffusion are subject of identification, while the optimal control acts on the drift function.

The paper objective consists of using the IPF specifics for finding the dynamic model of this stochastic system and the synthesis of the system optimal control function, which provides a joint system's identification with the optimal control's action. Such dynamic model reveals some regularities of a random process, while a randomness and uncertainty of the initial process could cover its regularities.

Conventional information science considers an information *process*, but traditionally uses the probability measure for the random *states* and Shannon's entropy measure as the uncertainty *function* of these states [34, 16, 19, 20, other].

The IPF controllable process' functional *information* measure can be applied as a given performance criterion, which might be specified for any particular performance criterion [5, 28, 35, others]. The information form of the found dynamic model allows encoding of the identified process' information measure, as well as each optimal control's action. These extend the approach applications to a wide class of controllable information systems.

Focusing in this paper on the approach essence, we omit the proofs of published results [22-28] and simplify the presentation. The paper is organized as follows:

Sec.2 introduces the entropy functional (EF) information measure for a *class* of random systems, modeled by the solutions of controlled Ito's stochastic differential equations.

In Sec.3 we evaluates the EF, using an operation of cutting-off the process functional, which is implemented by an impulse control function and delivered a hidden information of Feller's kernel.

In Sec.4 we formulate optimization problem for the EF, which determines the EF minimax principle and a variation problem (VP). The VP includes a *dynamic* approximation of the EF by corresponding information *path functional* (IPF) and finding both the diffusion process' information dynamic model (IDM) and the optimal control from the IPF extreme.

Sec.5 presents *specifics* of this VP *solution* using both Kolmogorov's (K) equation for the functional of a Markov process and the Jacobi-Hamilton (JH) equation for the IPF dynamics. The completion of both JH and K equations in the same field's region of a space becomes possible at some "punched" *discretely* selected points (DP), where JH imposes a dynamic constraint on the K solutions. As a result, the extremal trajectory, following from the VP, is divided on the extremal *segments*, each of which approximates a related segment of the random (microlevel) process (between the DP punched localities) with a *maximal probability*, as a segment of a *macroprocess*. At a DP exists a "*window*", where the random information affects the dynamic process on the extremals, creating its piece-wise *dependency* upon the observed data and the dynamic model's possibility of forming an optimal piece-wise control, applied to the diffusion process. These specifics allow us the *identification* of both Ito's and dynamic model's operators in *real time under* the optimal control action. The IPF dynamic process models the initial random process along its trajectories, measured by EF, with a maximal probability along these trajectories.

Sec.6 applies the VP results to find the identification equations and optimal control functions. The synthesized optimal control starts a stepwise action at the beginning of each segment and terminates this action at the segment end (where the identification takes place at the DP), and then connects the segments in the macrodynamic optimal process.

Information invariants (Secs.7-8), following from the VP, allow prediction of a discrete interval of the optimal control's action for each following extremal with a maximal probability, based on the process identification at each current DP.

In this optimal dual strategy, the concurrently synthesized optimal control allocates each future DP and provides the process identification jointly with the optimal control actions, which also filters the random process' between the DP.

The identified controllable dynamics for a multidimensional process (Sec.9) include *consolidation and aggregation* of the information macroprocess' trajectories in an information network's (IN) *structure* and ~~the~~ generation of the IN code.

Secs.10-11present the methodology for measuring, evaluation of the control's extracted information and procedure of dynamic modeling and prediction the identified diffusion process.

Sec.12. connects the IPF approach with Shannon's information theory, which leads to an optimal encoding of the identified dynamic operator, control and the IN by applying the model's information invariants for each Ito's *stochastic equation* with specific functions of the drift-vector and diffusion components. Attachments (A0, A1) focus on specifics of considered information process and its information measure, and (A2-A3) illustrates the paper results by Examples, which also demonstrate the procedure of problem solving in practical applications.

## *2. Information functional measure of a Markov diffusion process*

We consider a diffusion process, defined as a continuous Markov process, satisfying, generally, solutions of a $n$-dimensional controllable differential equations in Ito's form:

$$d\tilde{x}_t = a(t,\tilde{x}_t,u_t)dt + \sigma(t,\tilde{x}_t)d\xi_t, \tilde{x}_s = \eta, t \in [s,T] = \Delta, s \in [0,T] \subset R^1_+, \quad (2.1)$$

with the standard limitations [9, 21] on the functions of drift $a^u = a(t,\tilde{x}_t,u_t)$ (depending on control $u_t$), diffusion $\sigma = \sigma(t,\tilde{x}_t)$, and Wiener process $\xi_t = \xi(t,\omega)$, which are defined on a probability space of the elementary random events $\omega \in \Omega$ with the variables located in $R^n$; control $u_t$ is a piece-wise continuous function. (The drift and diffusion functions are defined through the process' probability and the solutions of (2.1)[8]).

Suppose that control function $u_t$ provides the transformation of a priory probability $P^a_{s,x}(d\omega)$ to a posteriori probability $P^p_{s,x}(d\omega)$, where a priory process $\tilde{x}^a_t$ is a solution of (2.1) prior to applying this control, at $a(t,\tilde{x}_t,u_t)=0$, and a posteriori process $\tilde{x}^p_t$ is a solution of (2.1) after such a control provides this transformation at $a(t,\tilde{x}_t,u_t)\neq 0$.

Such a priori process $\tilde{x}^a_t = \int_s^t \sigma(v,\zeta_v)d\zeta_v$ models an uncontrollable noise with $E[\tilde{x}^a_t]=\mathrm{O}$.

The process' $\tilde{x}^p_t$ transformed probability is defined through its transition probability

$$P^p(s,x,t,B)=\int_{x_t\in B}(p(\omega))^{-1}P^a_{s,x}(d\omega) \qquad (2.2)$$

with a probability density measure [32] (A0.1) of this transformation in the form

$$p(\omega)=\frac{P^a_{s,x}(d\omega)}{P^p_{s,x}(d\omega)}=\exp\{-\varphi^t_s(\omega)\}, \qquad (2.3)$$

which for the above solutions of (2.1) is determined through the additive functional of the diffusion process (Dynkin[18]):

$$\varphi^T_s = 1/2\int_s^T a^u(t,\tilde{x}_t)^T(2b(t,\tilde{x}_t))^{-1}a^u(t,\tilde{x}_t)dt-\int_s^T(\sigma(t,\tilde{x}_t))^{-1}a^u(t,\tilde{x}_t)d\xi(t). \qquad (2.4)$$

Using the definition of quantity information $I_{ap}$, obtained at this transformation, by entropy measure [A1]:

$$E_{s,x}\{-\ln[p(\omega)]\}=S_{ap}=I_{ap},\ S_{ap}=E_{s,x}\{\varphi^t_s(\omega)\} \qquad (2.5)$$

and after substituting the math expectation of (2.4) in (2.5) (at $E[\tilde{x}^a_t]=\mathrm{O}$) we get the entropy integral functional for a transformed process $\tilde{x}_t$ [25]:

$$S_{ap}[\tilde{x}_t]|_s^T=1/2E_{s,x}\{\int_s^T a^u(t,\tilde{x}_t)^T(2b(t,\tilde{x}_t))^{-1}a^u(t,\tilde{x}_t)dt\}=\int_{\tilde{x}(t)\in B}-\ln[p(\omega)]P^a_{s,x}(d\omega)=-E_{s,x}[\ln p(\omega)], \qquad (2.6)$$

where $a^u(t,\tilde{x}_t)=a(t,\tilde{x}_t,u_t)$ is a drift function, depending on control $u_t$, and $b(t,\tilde{x}_t)$ is a covariation function, describing its diffusion component in (2.1); $E_{s,x}$ is a conditional to the initial states ($s,x$) mathematical expectation, taken along the $\tilde{x}_t=\tilde{x}(t)$ trajectories.

Entropy functional (2.6) is an *information indicator* of a *distinction* between the processes $\tilde{x}^a_t$ and $\tilde{x}^p_t$ by these processes' measures; it measures a *quantity of information* of process $\tilde{x}^p_t$ regarding process $\tilde{x}^a_t$. For the process' equivalent measures, this quantity is zero, and it takes a positive value for the process' nonequivalent measures.

The definition (2.3) specifies *Radon-Nikodym's density* measure for a probability density measure, applied to entropy of a random process (Stratonovich [71]).

The *quantity of information* (2.5), is an equivalent of Kullback–Leibler's divergence (KL) for a continiuos random variables (Kullback [20]):

$$D_{KL}(P^a_{x,s}\,\|\,P^p_{x,s})=\int_X \ln(\frac{dP^a_{x,s}}{dP^p_{x,s}})dP^p_{x,s}=E_x[\ln\frac{P^a_{x,s}(\omega)}{P^p_{x,s}(\omega)}]=E_x[\ln p(\omega)]=-S_{ap}, \qquad (2.7)$$

where $\frac{dP^a}{dP^p}$ is Radon –Nikodym derivative of probability $P^a$ with respect to probability $P^p$:

$P^a(X)=\int\limits_X P^a_{x,s}(dx)\,, P^p(X)=\int\limits_X P^p_{x,s}(dx)$, while (2.7) determines a nonsymmetrical distance's measure between the entropies $S_a$ and $S_p$ related to these probabilities.

The KL measure is connected to both Shannon's conditional information and Bayesian inference (Jaynes [17]) of testing a priory hypothesis (probability distribution) by a posteriori observation's probability distribution.

Finally, definition of information integral *information measure of transformation*, applied to a *process'* probabilities, generalizes some other information measures.

## *3. The information evaluation of a Markov diffusion process by an entropy functional measure on the process' trajectories*

Advantage of the EF over Shannon's information measure consists in evaluating the inner connection and dependencies of the *random* process' states, produced at the generation of the process, which allows to measure the concealed information. Such a functional information measure is able to accumulates the process' information, *hidden* between the information states, and hence, brings more information then a sum of the Shannon's entropies counted for all process' separated states.

We introduce a method of cutting off the process on the separated states by applying an impulse control, which is aimed to show that cutting off the EF integral information measure on the separated states' measures decreases the quantity of process information by the amount which was concealed in the connections between the separate states. The $\delta$-cut-off of the diffusion process, considered below (sec.3a), allows us to evaluate the *quantity* of information which the functional EF conceals, while this functional binds the correlations between the non-cut process states. The cut-off leads to dissolving the correlation between the process cut-off points, losing the functional connections at these discrete points.

### 3a. *The step-wise and impulse controls' actions on functional* (2.6) *of diffusion process* $\tilde{x}_t$.

The considered control $u_t$ is defined as a piece-wise continuous function of $t\in\Delta$ having opposite stepwise actions:

$$u_+ \overset{def}{=} \lim_{t\to\tau_k+o} u(t,\tilde{x}_{\tau_k})\,, u_- \overset{def}{=} \lim_{t\to\tau_k-o} u(t,\tilde{x}_{\tau_k})\,, \qquad (3.1)$$

which is differentiable, excluding a set

$$\Delta^o=\Delta\setminus\{\tau_k\}_{k=1}^m\,, k=0,\ldots,m\,. \qquad (3.2)$$

The jump of the control function $u_-$ in (3.1) from a moment $\tau_{k-o}$ to $\tau_k$, acting on a *diffusion* process $\tilde{x}_t=\tilde{x}(t),$ might "cut off" this process after moment $\tau_{k-o}$.

The "cut off" diffusion process has the same drift vector and the diffusion matrix as the initial diffusion process.

Functional (2. 6), expressed via the process additive functional $\varphi_s^T:\Delta S[\tilde{x}_t]\,|_s^T=E_{s,x}[\varphi_s^T]$, for this "cut off" acquires a form ( Prochorov, Rozanov [32]):

$$\varphi_s^{t-}=\begin{cases}0, t\le\tau_{k-o};\\ \infty, t>\tau_k.\end{cases} \qquad (3.3)$$

The jump of the control function $u_+$ (3.1) from $\tau_k$ to $\tau_{k+o}$ might cut off the diffusion process *after* moment $\tau_k$ with the related additive functional

$$\varphi_s^{t+}=\begin{cases}\infty, t>\tau_k;\\ 0, t\le\tau_{k+o}.\end{cases}. \qquad \mathbf{(3.4)}$$

At the moment $\tau_k$, between the jump of control $u_-$ and the jump of control $u_+$, we consider a control *impulse*

$$\delta u^{\mp}_{\tau_k} = u_-(\tau_{k-o}) + u_+(\tau_{k+o}) . \qquad (3.5)$$

The related additive functional at a vicinity of $t = \tau_k$ acquires the form of an impulse function

$$\varphi_s^{t-} + \varphi_s^{t+} = \delta\varphi_s^{\mp} . \qquad (3.6)$$

The entropy functional at the localities of the control's switching moments (3.2) takes the values

$$S_- = E[\varphi_s^{t-}] = \begin{cases} 0, t \le \tau_{k-o}; \\ \infty, t > \tau_k . \end{cases} \text{ and } S_+ = E[\varphi_s^{t+}] = \begin{cases} \infty, t > \tau_k; \\ 0, t \le \tau_{k+o}. \end{cases}, \qquad (3.7)$$

changing from 0 to $\infty$ and back from $\infty$ to 0 and acquiring an *absolute maximum* at $t > \tau_k$, between $\tau_{k-o}$ and $\tau_{k+o}$. The related multiplicative functionals are

$$p_s^{t-} = \begin{cases} 0, t \le \tau_{k-o} \\ 1, t > \tau_k \end{cases} \text{ and } p_s^{t+} = \begin{cases} 1, t > \tau_k \\ 0, t \le \tau_{k+o} \end{cases}, \qquad (3.7a)$$

which determine probabilities $\tilde{P}_{s,x}(d\omega) = 0$ at $t \le \tau_{k-o}, t \le \tau_{k+o}$ and $\tilde{P}_{s,x}(d\omega) = P_{s,x}(d\omega)$ at $t > \tau_k$.

For the "cut-off" diffusion process, transitional probability (at $t \le \tau_{k-o}$, $t \le \tau_{k+o}$) turns to zero.

Then, the states $\tilde{x}(\tau - o), \tilde{x}(\tau + o)$ become independent, and the mutual time *correlations are dissolved*:

$$r_{\tau-o,\tau+o} = E[\tilde{x}(\tau - o)\tilde{x}(\tau + o)] \to 0 . \qquad (3.7b)$$

The entropy $\delta S_-^+(\tau_k)$ of the additive functional $\delta\varphi_s^{\mp}$, produced within, or at a border of the control impulse (3.5), is define by the equality

$$E[\varphi_s^{t-} + \varphi_s^{t+}] = E[\delta\varphi_s^{\mp}] = \int_{\tau_{k-o}}^{\tau_{k+o}} \delta\varphi_s^{\mp} P_\delta(d\omega) , \qquad (3.8)$$

where $P_\delta(d\omega)$ is a probability evaluation of impulse $\delta\varphi_s^{\mp}$.

Taking integral of the $\delta$-function $\delta\varphi_s^{\mp}$ between the above time intervals, we get at the border: $E[\delta\varphi_s^{\mp}] = 1/2 P_\delta(\tau_k)$ at $\tau_k = \tau_{k-o}$, or $\tau_k = \tau_{k+o}$. The impulse, produced by the controls, is a non random with $P_\delta(\tau_k)$ =1, which brings the EF estimation at $t = \tau_k$:

$$S^{\delta u}_{\tau_k} = E[\varphi_s^{\mp}] = 1/2 . \qquad (3.9)$$

This entropy increment evaluates an information contribution from the impulse controls (3.5) at a vicinity of the above discrete moments. Since that, each information contribution

$$E[\varphi_s^{t-}]_{\tau_k} = S^{u_-}_{\tau_k} \text{ and } E[\varphi_s^{t+}]_{\tau_k} = S^{u_+}_{\tau_k}$$

at a vicinity of $t = \tau_k$, produced by the corresponding controls' step functions $u_-(\tau_k), u_+(\tau_k)$ in (3.1), (3.5) accordingly, can be estimated by

$$S^{u_-}_{\tau_k} = 1/4,\ u_- = u_-(\tau_k),\ \tau_{k-o} \to \tau_k;\ S^{u_+}_{\tau_k} = 1/4,\ u_+ = u_+(\tau_k),\ \tau_k \to \tau_{k+o}, \qquad (3.10)$$

where the entropy, according to its definition (2.5), is measured in the units of Nat (1 Nat $\cong$ 1.44bits).

Estimations (3.9), (3.10) determine the entropy functional's cut-off values at the above time's borders under actions of these controls, which decreases the quantity of the functional's information by the amount that had been concealed before the cutting the process correlations (3.7b).
The EF definition through the *Radon-Nikodym's probability density* measure (2.3),(3.7a), *holds* also the considered cut-off peculiarities of the controllable process' EF.
Dissolving the correlation between the process cut off points leads to *losing the functional connections* at these discrete points, while the Feller's kernel measure of these connections and its relation to the additive functional's cutoff is *a source of a kernel information,* estimated by (2.12) through the impulse control actions. Moreover, since a jump action on Markov process, associated with "killing its drift", selects Feller's measure of the kernel [15a, other], considered cut-off of the information functional provides *information measure of the Feller kernel.*

***3b. The information estimation of the impulse control's cut-off action on diffusion process*** $\tilde{x}_t$ **.**

How much information is lost at these points? The evaluated information effect of losing the functional's bound information at these points holds the amount of 0.5 Nats (~0.772 bits) at each cut-off in the form of standard $\delta$-function; and *n* of a such cut-off loses information $I_c \cong 0.772n, n=1,2,3,...,$ .(3.11)

Thus, the process functional's information measure encloses $I_c$ bits more, compared to the information measure applied separately to each *n*-states of this process. The same result is applicable to a comparative information evaluation of the divided and undivided portions of an information process, measured by their corresponding EF. This means that an information process holds more information than any divided number of its parts, and the Entropy Functional measure of this process is also able to evaluate the quantity of information that *connects* these parts. As a result, the additive principle for the information of a process, measured by the EF, is *violated*:

$$\Delta S[\tilde{x}_t]|_s^T \geq \Delta S_1[\tilde{x}_t]|_s^{t_1} + \Delta S_2[\tilde{x}_t]|_{t_1+o}^{t_2} + \Delta S_m[\tilde{x}_t]|_{t_2+o}^{t_m} + \Delta S_T[\tilde{x}_t]|_{t_{m+o}}^{T},..., \quad (3.11a)$$

where the process $\tilde{x}_t$ is cutting-off at the moments $t_1, t_1+o;....t_m, t_{m+o};....$

Therefore integral functional's measure accumulates more process information than the sum of the information measures in its separated states.
This entropy increments evaluate an information contribution from the impulse controls at a vicinity of the above discrete moments. Each of the impulse control could be produced by two corresponding controls' step functions: the left stepwise control acting down, and the right stepwise control acting up $u_-(\tau_k), u_+(\tau_k)$ .
Acting together, both of them extract these information contributions producing the process' cutt-off.
While the above estimations determine the entropy functional's cut-off values at the above time intervals under actions of these controls, which decreases the quantity of the process' functional information by the amount that had been concealed before cutting the process correlations.

***4. An optimal information transformation, its information measure and a minimax principle***

Let us have information measure (2.6) for diffusion process (2.1) , and *find* the condition of its optimization in the form

$$\min S[\tilde{x}_t]|_s^T = S[\tilde{x}_t^o]|_s^T \quad , \quad (4.1)$$

where $\tilde{x}_t^o$ is an extremal trajectory minimizing this functional, and $S[\tilde{x}_t^o]|_s^T$ is a functional minimizing $S[\tilde{x}_t]|_s^T$ .

<u>Proposition 4.1.</u> Minimization of information transformation $E_{s,x}[-\ln p[\tilde{x}_t]]$ with $p = P^a_{s,x} / P^p_{s,x}$, where $P^a_{s,x}$ is differential probability of a priory process $\tilde{x}^a_t$, transformed to differential probability $P^p_{s,x}$ of *a posteriori* process $\tilde{x}^p_t$, leads to *converting* the probability measure $p$ in optimal $p_o = P^{ap}_{s,x} / P^{pd}_{s,x}$, which measures the transformation of random a priori trajectories $\tilde{x}^{ap}_t$ (with differential probability $P^{ap}_{s,x}$ ) to an *optimal predictable information process* $\tilde{x}^o_t \to x^{pd}_t$ (with probability measure $P^p_{s,x} \to 1$). (Integration of each differential probability determines the corresponding transitional probability distributions $P(s,x,t,B)$ according to (2.2)).

*Proof.* At $S[\tilde{x}_t]|^T_s = S_{ap}$, we get $\min S_{ap} = \min - E_x[\ln(P^a_{s,x} / P^p_{s,x})] = \min(S_a - S_p)$, (4.2)

where $S_a = E_{s,x}[-\ln(P^a_{s,x})], S_p = E_{s,x}[-\ln(P^p_{s,x})]$ are the entropy functionals of related processes.

At $\min S_{ap} = S^o_{ap} = (S_{ao} - S_{pd})$ we have $(S_{ao} - S_{pd}) = \min(S_a - S_p)$, (4.2a)

where *a posteriori dynamic* process $x^{pd}_t$ with a differential probability $P^{pd}_{s,x} \to 1$ is a *predictable* process.

For such process

$$S_{pd} = -E_{s,x}[\ln P^{pd}_{s,x}] = -\ln P^{pd}_{s,x} = -\ln(P^{pd}_{s,x} \to 1) \to 0, \quad (4.3)$$

and (4.2a) takes the form

$$\min(S_a - S_p) = \min S_a - \min S_p = \min S_a = \min S_{ao} = S^o_{ap}, \quad (4.3a)$$

*satisfying to a minimum for information* transformation (4.1).

This means, such *a posteriori dynamic* process $x^{pd}_t$ is an *optimal process* minimizing (4.1), as its *extremal*, and from (4.3) it follows that $x^{pd}_t$ is the *optimal predictable* process.•

<u>Comment 4.1.</u>

(1). A random priori process $\tilde{x}_t$, holding an entropy functional measure (4.1): $S_{ap}[\tilde{x}_t]$, at satisfaction of (4.3a), is measured by the optimal measure of a priori process in the form :

$$S_{ao}[\tilde{x}^{ap}_t] = \min S_{ap}[\tilde{x}_t] = S^o_{ap}. \quad (4.4).$$

(2).The minimum in (4.2) not necessarily leads to (4.3a): all other transformed processes with $\min S_p > 0,$ have

$$\min(S_a - S_p) < S^o_{ap}. \quad (4.4a)$$ •

<u>Corollary 4.1.</u>

Optimal transformation (4.1) , which depends only on the entropy *minimum* for a priori process (4.4), has a *maximum of minimal entropy* $S_{ap}$ (4.1) among other transformed processes, therefore, it satisfies the *entropy's minimax principle* :

$$\max\min(S_a - S_p) = \min S_{ap} = S^o_{ap}, \quad (4.5)$$

which we write in information form at $I_{ap} = S_{ap}, \max P^p_{s,x} \to 1$:

$$\max\min - E_{s,x}[\ln(P^a_{s,x} / P^p_{s,x})] = \max\min E_{s,x}[-\ln p(\omega)]] = \max\min I_{ap}. \quad (4.6)$$

Hence, the sought dynamic process will be reached through such a minimal entropy transformation, which passes the minimum entropy of a priory process to a posteriori process with a maximum probability.

This means, the maximum of a minimal information provides such information transformation a random process to a dynamic process, which approximates the random process with a maximal probability.

Otherwise: the *MiniMax information principle* applied to a random process, implements optimal transformation (4.1) via an extraction of the process' information regularities with a maximal probability.•

Process $x_t$ gets dynamic properties, by holding a nonrandom functional $S\,[x_t\,]|_s^T$ , which we call *information path functional* (IPF) on trajectories of its extremals $x_t$ . The IPF information measure on extremals $x_t \to x_t^{pd}$ approximates random trajectories *of a priori process* $\tilde{x}_t \to \tilde{x}_t^{ap}$ with a *maximal* probability functional measure.

In this optimal approximation, a posteriori dynamic process brings a macroscopic evaluation of the priory random process, defining its *macrodynamic* process.

The IPF is a dynamic analogy of the EF, and the dynamic trajectories $x_t$ model diffusion process $\tilde{x}_t$ .

The known *Maximum Entropy* (Information) principle, applied directly to a *random process*, would bring a related *MaxEnt* optimal process, having a *minimal probability distribution* (among all non *MaxEnt* optimal process), thereafter being the most *uncertain to indentify*.

While maximum entropy is associated with disordering and a complexity, minimum entropy means ordering and simplicity. That is why extracting a maximum of minimal information means also obtaining this maximum from a most ordered transformation. To retain regularities of the transformed process, such transformation should not lose them, having a *minimal* entropy of the random process. The optimal transformation spends a minimal entropy (of a lost) for transforming each priory random process to its dynamic model.

Hence, the transformed process that holds regularities should satisfy some *variation principle*, which according to R. Feynman [13], could be applied to a process as *a mathematical form of a law*, expressing the process regularities.

We formulate this law through imposing the information *MiniMax* variation principle (VP) on the random process, which brings both its optimal transformation, minimizing the entropy for a random dynamic process and maximizes this minimum. Solving the VP allows us finding the IPF functional $S\,[x_t\,]|_s^T$ and its extremals $x_t$ , which approximate a posteriori random process with a maximum probability on its dynamic trajectories.

Thus, *MiniMax* principle is implemented by the VP through minimization of entropy functional (EF) of random process, whose minimum is maximized by information path functional (IPF) of information macrodynamics.

The EF-IPF approach converts the *uncertainty* of a random process into the *certainty* of a dynamic information process.

### ***5. The essence of the information path functional approach***

The considered variation problem (VP) in the form

$$\min_{\tilde{x}_t(u)} \tilde{S}[\tilde{x}_t] = \min_{x_t(u)} S[x_t], \qquad (5.1)$$

connects the entropy functional (2.6) (defined on a random trajectories) with the *IPF integral functional*

$$S = \int_s^T L(t,x,\dot{x})dt = S[x_t]. \qquad (5.1a)$$

<u>*Proposition*</u> <u>5.1.</u> The *solution* of variation problem (5.1) for the entropy functional brings the following equations of extremals for a vector $x$ and a conjugate vector $X$ accordingly:

$$\dot{x} = a^u\,,\ (t,x) \in Q\ , \qquad (5.2)$$

$$X = (2b)^{-1} a^u\,; \qquad (5.3)$$

and the constraint

$$a^u(\tau)X(\tau) + b(\tau)\frac{\partial X}{\partial x}(\tau) = 0\,, \qquad (5.4)$$

imposed on the solutions (5.2), (5.3) at some set

$$Q^o = R^n \times \Delta^o, \Delta^o = [0,\tau], \tau = \{\tau_k\}, k = 1,...,m\ , Q^o \subset Q\,, \qquad (5.4a)$$

where controls (3.8), (3.9), applied at the discrete moments $(\tau_k)$ (5.4a), implement these VP solutions. •

*Proof*. Using the Jacobi-Hamilton (JH) equation[10] for a function of action $S = S(t,x)$, defined on the extremals $x_t = x(t)$ of functional $S[x_t]$ at $(t,x) \in Q$, we have

$$-\frac{\partial S}{\partial t} = H, H = \dot{x}^T X - L, \quad (5.5)$$

where $X$ is a conjugate vector for $x$ and $H$ is a Hamiltonian for this functional.
(All derivations here and below have vector form).
Let us consider the distribution of functional (2.6) on $(t,x) \in Q$ as a function of current variables $\tilde{S} = \tilde{S}(t,x)$, which satisfies the Kolmogorov (K) equation [8, 32, others], applied to the math expectation of functional (2.6) in the form:

$$-\frac{\partial \tilde{S}}{\partial t} = (a^u)^T \frac{\partial \tilde{S}}{\partial x} + b\frac{\partial^2 \tilde{S}}{\partial x^2} + 1/2(a^u)^T(2b)^{-1}a^u \ . \quad (5.5a)$$

From condition (5.1) it follows

$$\frac{\partial S}{\partial t} = \frac{\partial \tilde{S}}{\partial t}, \frac{\partial \tilde{S}}{\partial x} = \frac{\partial S}{\partial x} \quad (5.5b)$$

where $\frac{\partial S}{\partial x} = X, -\frac{\partial S}{\partial t} = H$ .
This allows us to join (5.5a) and (5.5b) in the form

$$-\frac{\partial \tilde{S}}{\partial t} = (a^u)^T X + b\frac{\partial X}{\partial x} + 1/2a^u(2b)^{-1}a^u = -\frac{\partial S}{\partial t} = H \ . \quad (5.6)$$

Applying to (5.6) the Hamilton equation $\frac{\partial H}{\partial X} = \dot{x}$ we get (5.2), and after substituting it to (5.5) (at the fulfillment of (5.5a)) we come to Lagrangian

$$L = -b\frac{\partial X}{\partial x} - 1/2\dot{x}^T(2b)^{-1}\dot{x} \ . \quad (5.6b)$$

Here function $a^u = a^u(t,x_t)$, defined on extremals $x_t$, is a dynamic analogy of the drift function $a^u = a(t,\tilde{x}_t,u_t)$ in Eqs (2.1), and $b = b(t,x_t)$ is a dynamic analogy of the dispersion matrix $b(t,\tilde{x}_t)$ in (2.1).
After substitution of both nonrandom $a^u$ and $b$ to the conditional math expectation (2.6) we get the integral functional $\tilde{S}$ on the extremals:

$$\tilde{S}_e[x(t)] = 1/2\int_s^T (a^u)^T(2b)^{-1}a^u dt = S[x_t] \ . \quad (5.7)$$

which should satisfy the variation condition (5.1), or

$$\tilde{S}_e[x(t)] = S[x(t)] , \quad (5.7a)$$

where both integrals are determined on the same extremals.
From (5.7a) it follows

$$L = 1/2(a^u)^T(2b)^{-1}a^u , \text{ or } L = 1/2\dot{x}^T(2b)^{-1}\dot{x} \ . \quad (5.7b)$$

Both expressions for Lagrangian in the forms (5.6b) and (5.7b) coincide at

$$b\frac{\partial X}{\partial x} + \dot{x}^T(2b)^{-1}\dot{x} = 0 \ . \quad (5.7c)$$

Applying to (5.7b) (at the fulfillment of (5.2)) the Lagrange's equation $\frac{\partial L}{\partial \dot{x}} = X$, we get the equations for the conjugate vector $X = \dot{x}^T(2b)^{-1}$ and $\dot{x} = L_k X, L_k = 2b$, which prove (5.3).

The fulfillment of (5.6) is possible if equation (5.7c) is satisfied at each point $(t,x)\in Q$ of the functional field $\tilde{S}(x(t))$, except a certain set (5.4):

$$Q^o \subset Q,\ Q^o = R^n \times \Delta^o, \Delta^o = [0,\tau], \tau = \{\tau_k\}, k = 1,...,m ;$$

where the following relations hold true:

$$E_{x,\tau}[(a^u)^T \frac{\partial \tilde{S}}{\partial x} + b\frac{\partial^2 \tilde{S}}{\partial x^2}] = 0, a^u = a^u(t,x), b = b(t,x) \ . \qquad (5.7d)$$

Substituting to (5.7d) the relations (5.7) and (5.1) we come to (5.4).
This equation determines the dynamic constraint, which JH imposes on K to fulfill (5.1). •

Corollary 5.1.

The control action on equation (5.6), which implements the variation conditions (5.1) at the set of discrete moments (5.4a**)**, requires *turning* the constraint (5.4) *on* with changing the process' function $-\partial\tilde{S}/\partial t$ from its *maximum* to a *minimum.*

*Proof*. At the satisfaction of (5.4), Hamiltonian (5.6) acquires the form

$$H = 1/2(a^u)^T(2b)^{-1}a^u \ , \qquad (5.8)$$

which, after substituting (5.2) to (5.6), corresponds to a $\min(-\partial\tilde{S}/\partial t) = -\partial\tilde{S}_o/\partial t$ (of this function on the extremals when it coincides with the Hamiltonian):

$$\min_{x(t)}(-\partial\tilde{S}/\partial t) = H = 1/2(\dot{x})^T(2b)^{-1}\dot{x} \ , \qquad (5.8a)$$

whereas function $(-\partial\tilde{S}/\partial t)$ in (5.5a) reaches its *maximum* when the constraint is not imposed.

Both minimum and maximum are conditional with respect to the constraint imposition. •

Corollary 5.2.

(1).While both Hamilton Eqs (with Hamilatonians in forms (5.6) and (5.8)) accordingly) provide *extremal* solutions *minimizing* functional (5.1), the extremal with Hamiltonian (5.6) minimizes this functional with a *maximal* speed $|\partial\tilde{S}/\partial t| > |\partial\tilde{S}_o/\partial t|$ compared to the extremal with Hamiltonian (5.6).

(2).These maximal extremal's solutions approximate a priori diffusion process with a minimum of maximal probability, satisfying $\max|\partial\tilde{S}/\partial t|$, and a maximum of the probability, satisfying $\min|\partial\tilde{S}/\partial t| = |\partial\tilde{S}_o/\partial t|$ accordingly.•

Solution of this variation problem's (VP) [24] (with the detailed proofs in Lerner [23]) *automatically* brings the *constraint*, imposing *discretely* at the states' set (5.4a) by the applied optimal controls (synthesized in Sec.6), which change the entropy derivation from its maximum to its minimum.

This means the *discrete control action* is provided by the VP though imposing the constraint.

Consequently, an extremal is determined by imposing such dynamic constraint during the control actions, which select the extremal *segment* from the initial random process (Fig.1a).•

Below we find the limitations on completion of constraint equation (5.4), which also restrict the controls action, specifically when it should be turn off.

Proposition 5.2.

Let us consider diffusion process $\tilde{x}(s,t)$ at a locality of states $x(\tau_k - o)$, $x(\tau_k)$, $x(\tau_k + o)$, formed by the impulse control action (Sec.3a), where the process is cutting off *after* each moment $t \le \tau_k - o$ *-at* $t > \tau_k$, and each moment $t \ge \tau_k + o$ is *following* to the cut-off , with $(\tau_k - o) < \tau_k < (\tau_k + o)$.

The additive and multiplicative functionals (Sec.3a) satisfy Eqs (3.7,3.7a) at these moments.

Then the constraint (5.7d) acquires the form of operator $\tilde{\tilde{L}}$ in Eq.

$$-\frac{\partial \Delta\tilde{S}}{\partial s}=\tilde{\tilde{L}}\Delta\tilde{S}\,,\Delta\tilde{S}(s,t)=\begin{cases}0,t\le\tau_k-o;\\ \infty,t>\tau_k;\end{cases} \qquad (5.9)$$

which at $\Delta\tilde{S}(s,t\le\tau_k-o)=0$ (5.9a) satisfies Eq

$$\tilde{\tilde{L}}=(a^u)^T\frac{\partial}{\partial x}+b\frac{\partial^2}{\partial x^2}=0. \qquad (5.9b)$$

The *proof* follows from [32], where it is shown that relation $\Delta\tilde{S}(s,t\le\tau_k-o)=E_{s,x}[\varphi_s^{t-}]$ satisfies to operator $\tilde{\tilde{L}}$ in Eq (5.9), which is connected with operator $\tilde{L}$ of the initial K Eqs (5.5a) by relation $\tilde{\tilde{L}}=\tilde{L}-1/2(a^u)^T(2b)^{-1}a^u$ .
From these relations, at completion of (5.9a), we get (5.9b) and then

$$E_{s,x}[(a^u)^T\frac{\partial\Delta\tilde{S}}{\partial x}+b\frac{\partial^2\Delta\tilde{S}}{\partial x^2}]=0, \qquad (5.9c)$$

where $\Delta\tilde{S}(s,t\le\tau_k-o)=E_{s,x}[\varphi_s^{t-}]=S_-$ is the process' functional, taken **prior to** the moment of cutting-off, when constraint (5.8c,5.7d) is still imposed. •
From [32] it also follows that solutions of (5.9c) allow classifying the states $x(\tau)=\{x(\tau_k)\},k=1,...,m$, considered to be the *boundary* points of a diffusion process **at** $\lim_{t\to(\tau_k+o)}\tilde{x}(t)\to x(\tau_k+o),(\tau_k-o)\le t\le(\tau_k+o)$ .
A boundary point $x(\tau_k+o)$ (that corresponds applying step-up control $u_+(\tau_k)$ (3.5) and sharp growing up the drift function $a^u=a(t,\tilde{x}_t,u_t)$ in (2.1)) *attracts* only if the function

$$R(x)=\exp\{-\int_{x_o}^{x}a^u(y)b^{-1}(y)dy\}, \qquad (5.9d)$$

defining the general solutions of (5.9c), is integrable at a locality **of** $x=x(\tau_k+o)$, satisfying the condition

$$|\int_{x_o}^{x_\tau}R(x)dx|<\infty\,. \qquad (5.9e)$$

A boundary point *repels* if (5.9e) does not have the limited solutions at this locality**;** it means that the above Eq. (5.9c) is not integrable in this locality.• The boundary dynamic states carry *hidden dynamic* connections between the process' states.
*Comments* 5.1.
(1). The constraint, imposed at $t\le(\tau_k-o)$, corresponds to the VP action *on the macrolevel*, whereas this action produces the cut-off at the following moment $\tau_k>(\tau_k-o)$ on the *microlevel* (both random and quantum [27,28]), when the constraints is turning off. Hence, the moment of imposing constraint (5.4) on the dynamic macrolevel coincides with the moment of imposing constraint (5.9b) on the microlevel, and the same holds true for the coinciding moments of turning both constraints off, which happen simultaneously with applying step-down control $u_-(\tau_k-o)$ (as a left part of the impulse control (3.5) that cuts-off the random process).
The state $x(\tau_k-o)$, prior to the cutting-off, has a minimal entropy $S[x(\tau_k-o)]\to 0$, which with a maximal probability puts it close to the entropy of a nonrandom dynamic macrostate on the extremal at the same moment of imposing the constraint.
Under the $u_-(\tau_k-o)$ control action (Sec.3a), the state $x(\tau_k-o)$ with a minimal entropy moves to a random state $x(\tau_k)$ with a maximal entropy, which turns the dynamics to a randomness.

(2).Under the $u_+(\tau_k + o)$ control action, state $x(\tau_k)$, with a maximal entropy $S[x(\tau_k)] \to \infty$ (according to (3.7)), moves to state $x(\tau_k + o)$, which, *after* the cutting-off, has a minimal entropy $S[x(\tau_k + o)] \to 0$ *as it approaches to dynamic state* $x(\tau^1_{k+1})$. This means that the state, currying a maximal information, transfers it to the state with a minimal information (and a maximal probability), which gets it with this maximal probability (Sec 4).
The state $x(\tau_k + o)$, having a maximal probability, could be transferred to a dynamic process, as starting point of the following extremal). Moreover, this state absorbs an increment of the entropy functional Sec.3a, (3.9a),(3.10) at the moment $\tau_k + o : S[x(\tau_k + o)] \cong 1/4 Nats$, which, with the state $x(\tau_k + o)$ is transported to a starting extremal segment of the macrodynamics, as its primary entropy contribution from the control $u_+(\tau_k + o)$.
The same way, the control $u_-(\tau_k - o)$ action had transferred the increment of the entropy functional (3.10) at the moment $\tau_k - o : S[x(\tau_k - o)] \cong 1/4 Nats$ from the ending point of a previous extremal segment to the diffusion process. This means, both ending and starting points of each extremal's segment possess the same two parts (3.9) of a total information contribution $S[u(\delta\tau_k)] \cong 1/2 Nats$, getting it from diffusion process.
This "hidden information", obtained from both bound dynamic connections between the process' states and the cut-off information contribution is a source of the information macrodynamics.
(3). A locality $x(\tau_k)$ of its border states $((x(\tau_k - o), x(\tau_k + o))$ forms "*punched*" discrete *points* (DP) $(..., x(\tau_{k-1}), x(\tau_k), x(\tau_{k+1}), ...)$ of the space $Q^o = R^n \times \Delta^o$, which establish a link between the microlevel's diffusion and macrolevel's dynamics. The macroequations (5.2, 5.3) act along each extremal, only approaching these points to get information from the diffusion process with the aid of an optimal control. The impulse control intervenes in a stochastic process (like that in [39]), extracts information (from the above locality) and transfers it to macrodynamics. The information contribution (in this locality) from the entropy functional (EF) $S_k[\delta\tilde{x}(\tau_k)]$ coincides with that for the information path functional (IPF) on each extremal segment's ending and starting **points** $I^k_s[\delta(\tau_k)]$ at $\delta\tau_k = \tau^1_{k+1} - \tau^o_k$.
The constraint equation is the *main mathematical structure* that distinguishes the equations of diffusion stochastics from related *dynamic* equations; at imposing the constraint, both equation's solutions coincide (at the process' border states).
(4). The optimal control, which implements VP by imposing the constraint (in the form (5.9b)) on the microlevel and (in the form (5.4)) on the macrolevel, *and* generating the macrolevel's dynamics, should have a *dual and simultaneous action* on both diffusion process and its dynamic model.
It's seen that the impulse control (IC), composed by two step-controls SP1, SP2 (forming the IC's left and right sides accordingly, Sec.3a, Fig.1b)), which provides the cut-off of the diffusion process, can turn *on* and *off* the constraint on the microlevel, and in addition to that, transfers the state with maximal probability, produced after the cut-off, to start the macrolevel's dynamics with initial entropy contribution, gaining from the cut-off.
The IC, performing the above dual operations, implements the VP, as its *optimal control*, which extracts most probable states from a random process, in the form of $\delta$-probability distribution.
We *specify* the following control's functions:
– The IC, applied to diffusion process, provides a sharp maximum of the process information, which coincides with the information of starting at the same moment the model's dynamic process. That implements the principle of maximal entropy at this moment;
– The start of the model's dynamic process, is associated with imposing the dynamic constraint (DC) on the equation for the diffusion process with the SP2 dual action, which also keeps holding along the extremal;

– At the moment when the DC finishes imposing the constraint, the SP2 stops, while under its stepwise-down action (corresponding to control SP1), the dynamics process on the extremal meets the punched (‘‘bounded’’) locality of diffusion process;
– At this locality, the IC is applied (with both SP1, stopping the DC, and the SP2, starting the DC at the next extremal segment), all processes sequentially repeat themselves, concurrently with the time course of the diffusion process;
– The starting impulse control binds the process’ maximum probability state, with the following start of the extremal movement, which keeps the maximum probability during its dynamic movement.
(5). Because the dynamic process depends on the moments’ of controls’ actions, which also depend on the information getting from the random process, a probability on trajectories of the dynamic process is less than 1.
Such dynamic process is *not a deterministic process*. Even though the dynamic process would have only a single initial random conditions and/or a single extremal segment, still it will be not a deterministic process.
(6). Since both the constraint imposition and the control action are limited by punched points $x(\tau_k), x(\tau_{k+1})$, the control, starting an extremal movement at the moment $\tau_k^o = \tau_k + o$, should be *turned off* at a moment $\tau_k^1 = \tau_{k+1} - o$ preceding to the following punched point $\tau_{k+1}$. And to continue the extremal movement, the control should be turned *on* again at the moment $\tau_{k+1}^o = \tau_{k+1} + o$ following $\tau_{k+1}$ to start a next extremal segment. This determines a *discrete* action of the stepwise control *during each interval of the extremal movement* $t_k = \tau_k^1 - \tau_k^o, t_{k+1} = \tau_{k+1}^1 - \tau_{k+1}^o, k = 1,...,m$, where $m$ is the number of applied controls. Both the control’s amplitude and time intervals will be found in Sec.6.
The constraint limits both time length and path of the extremal segment between the punched localities.
(A very first control is applied at the moment $\tau_o^o = s + o$, following the process Eqs’ initial condition at $t = s$ .)
The process continuation requires connecting the extremal segments between the punched localities by the joint stepwise control action: with a $k$ - control SP1 *turning off* at the moment $\tau_k^1$ while transferring to locality $\tau_{k+1}$, and a next $k+1$-control SP2, which, while transferring from $\tau_{k+1}$ locality, is *turning on* the following extremal segment at the moment $\tau_{k+1}^o$ .These controls provides a *feedback* from a *random process to macroprocess* and then *back* to a random process. By imposing constraints (5.4) and (5.9b) during each time interval $t_k$, both controllable random process (as an object) and its dynamic model become equal probable, being *prepared* (during this time delay) for getting new information and the optimal control action utilizing this information. Both stepwise controls form an impulse IC control function $\delta(x(\tau_k^1), x(\tau_{k+1}), x(\tau_{k+1}^o))$ acting between moments $\tau_k^1, \tau_{k+1}, \tau_{k+1}^o$ (or $\tau_{k-1}^1, \tau_k, \tau_k^o$) and implementing relation (3.5), which brings the EF peculiarities (3.7) at these moments, making the impulse control (3.5) a part of the VP implementation. Such a control imposes the constraint in the form (5.9b) on the initial random process, selecting its border points, as well as it starts and terminates the extremal movement between these points, which models each segment of the random process by the segment’s dynamics.
(7). The constraint in the form (5.4), allows generation of the process’ dynamics, following from the variation Eqs (5.1,5.2). Conditions (5.8) , (5.8a) determine the increments of functional (5.6) between the punched points during each interval $t_k$ of extremal movement, which preserves this increment:

$$\Delta S_c[x(t_k)] = inv,\ k = 1,...,m . \quad (5.10)$$

Equality (5.10) expresses the VP information invariant that depends on both functions $a^u, b$ in (2.6) and (5.8).
Condition (5.10) will be used to find both local information invariants (Secs.7,8) and interval of applying control $t_k$ .
(8).Generally, Hamiltonian Eqs of $n$-dimensional system lead to $n$ equations of a second order, with complex conjugated solutions, which are reversible in time (being invariant at changing its time course $t$ on $-t$ ). The initial Ito Eq (2.1) supposedly has the same dimension with their *irreversible* solutions, accessible at the punched localities.

***6. The optimal control synthesis, model identification, and specifics of both DC and IPF solutions***

Writing equation of extremals $\dot{x} = a^u$ in a dynamic model's traditional form (Alekseev *et al* [3]):

$$\dot{x} = Ax + u, u = Av, \dot{x} = A(x+v), \quad (6.1)$$

where $v$ is a control reduced to the state vector $x$, we find optimal control $v$ that solves the initial variation problem (VP) and identifies matrix $A$ under this control's action.

Proposition 6.1.

The reduced control is formed by a feedback function of macrostates $x(\tau) = \{x(\tau_k)\}, k = 1,...,m$:

$$v(\tau) = -2x(\tau), \quad (6.2)$$

or

$$u(\tau) = -2Ax(\tau) = -2\dot{x}(\tau), \quad (6.2a)$$

at the DP localities of moments $\tau = (\tau_k)$ (5.9), and the matrix $A$ is identified by the equation

$$A(\tau) = -b(\tau)r_v^{-1}(\tau), r_v = E[(x+v)(x+v)^T], b = 1/2\dot{r}, r = E[\tilde{x}\tilde{x}^T] = r_v(\tau) \quad (6.3)$$

through the above correlation functions, or directly, via the dispersion matrix $b$ from (2.1):

$$|A(\tau)| = b(\tau)(2\int_{\tau-o}^{\tau} b(t)dt)^{-1} > 0, \ \tau - o = (\tau_k - o), k = 1...,m. \quad (6.3a)\bullet$$

*Proof.* Using Eq. for the conjugate vector (5.3), we write the constraint (5.4) in the form

$$\frac{\partial X}{\partial x}(\tau) = -2XX^T(\tau), \quad (6.4)$$

where for model (6.1) we have

$$X = (2b)^{-1}A(x+v), X^T = (x+v)^T A^T(2b)^{-1}, \frac{\partial X}{\partial x} = (2b)^{-1}A, \ b \neq 0, \quad (6.4a)$$

and (6.4) acquires the form

$$(2b)^{-1}A = -2E[(2b)^{-1}A(x+v)(x+v)^T A^T(2b)^{-1}], \quad (6.4b)$$

for which at a nonrandom $A$ and $E[b] = b$, the identification equations (6.3) follow up. The completion of (6.3) and (6.4) is reached with the aid of the control's action, which we find using (6.4a) in the form

$$A(\tau)E[(x(\tau)+v(\tau))(x(\tau)+v(\tau))^T] = -E[\dot{x}(\tau)x(\tau)^T], \text{at } \dot{r} = 2E[\dot{x}(\tau)x(\tau)^T]. \quad (6.5)$$

This relation after substituting (6.1) leads to

$$A(\tau)E[(x(\tau)+v(\tau))(x(\tau)+v(\tau))^T] = -A(\tau)E[(x(\tau)+v(\tau))x(\tau)^T], \quad (6.5a)$$

and then to

$$E[(x(\tau)+v(\tau))(x(\tau)+v(\tau))^T + (x(\tau)+v(\tau))x(\tau)^T] = 0, \quad (6.5b)$$

which is satisfied at applying the control (6.2). The controlling $\dot{x}(\tau)$ determines control (6.2a) directly from (6.2a).

Since $x(\tau)$ is a discrete set of states, satisfying (5.4a), (5.8), the control has a discrete form, applied at this set. Each stepwise control (6.2) with its inverse amplitude -2 $x(\tau)$, doubling the controlled state $x(\tau)$, is applied at beginning of each extremal segment and acts during the segment's time interval. This control imposes the constraint (in the form (6.4), (6.4b)), which follows from the variation conditions (5.1, 5.1a), and, therefore, it implements this condition.

The same control, applied to both random process and the extremal segment, transforms $\tilde{x}_t^a$ to $\tilde{x}_t^p = \tilde{x}_t^{ao}$, and then, by imposing the constraint, transforms $\tilde{x}_t^{ao}$ to $\tilde{x}_t^{pd}$ extremals. During the last transformation, this control also initiates the identification of matrix $A(\tau)$ following (6.3,6.3a). Two of these stepwise controls, performing both of these transformations (which starts and terminates the constraint on each segment), form together a *single impulse control*, whose jump-down and jump-up actions at each segment's punched locality extract its hidden information.

Finding this control here just *simplifies* some results of Theorem [26]. •

Corollary 6.1. The control, turning the constraint on, creates a Hamilton dynamic model with complex conjugated eigenvalues of matrix $A$. After the constraint's termination (at the diffusion process' boundary point), the control transforms this matrix to its *real* form (on the boundary), which is identified by diffusion matrix in (6.3). •

Proposition 6.2.

(1)-The model (6.1) in the form of a closed system:

$$\dot{x}(t) = A^v x(t) \quad (6.6)$$

with a feedback control (6.2), leads to the same form of a drift vector for both models (2.1),(6.1):

$$a^u = A(\tau)(x(\tau)+v(\tau)) = A^v(\tau)x(\tau), \quad (6.6a)$$

which under control $v(\tau_k^o) = -2x(\tau_k^o)$, applied during time interval $t_k = \tau_k^1 - \tau_k^o$, by the moment $\tau_k^1$, gets the form

$$A^v(\tau_k^1) = -A(\tau_k^o)\exp(A(\tau_k^o)\tau_k^1)[2-\exp(A(\tau_k^o)\tau_k^1)]^{-1}. \quad (6.6b)$$

And the controllable dynamics is described by eigenfunctions $\lambda_i^v(t_i,\tau_k)_{i,k=1}^{n,m}$ of model's $A^v(t,\tau)$ operator.

(2)-At the DP localities of moments $\tau = (\tau_k, \tau_{k+1})$ (5.9), when controls $v(\tau_k^1) = -2x(\tau_k^1)$ and $v(\tau_{k+1}^1) = -2x(\tau_{k+1}^1)$ are applied, both matrixes change their signs according to relations

$$A^v[v(\tau_k^1)] = -A[v(\tau_k^1)], A^v(\tau_{k+1})_{v=0} = A(\tau_{k+1})_{v=0}, A^v[v(\tau_{k+1}^o)] = -A(\tau_{k+1})_{v=0}. \quad (6.6c)$$

(3)-At the moment of turning control $v_o(\tau) = -2x(\tau)$ to $v_o(\tau+o) = -2x(\tau+o) \to 0$, function $a^u(v_o) \to 0$.

*Proof* (1).Using both (6. 1) and (6.6):

$$a^u(\tau, x(\tau,t)) = A(\tau,t)(x(\tau,t)+v(\tau)); A(\tau)(x(\tau)+v(\tau)) = A^v(\tau)x(\tau) \quad (6.7)$$

at applying control (6.2), we get

$$A^v(\tau) = -A(\tau) = 1/2b(\tau)r_v^{-1}(\tau), b(\tau) = 1/2\dot{r}_v(\tau). \quad (6.7a)$$

In particular, by applying control $v(\tau_k^o) = -2x(\tau_k^o)$, which imposes the constraint during time interval $t_k = \tau_k^1 - \tau_k^o$ (starting it at $\tau_k^o$ and terminating at $\tau_k^1$) to both (6.1) and (6.6), we get their solutions by the *end* of this interval:

$$x(\tau_k^1) = x(\tau_k^o)[2-\exp(A(\tau_k^o)\tau_k^1)]. \quad (6.7b)$$

Substituting this solution to $\dot{x}(\tau_k^1) = A^v(\tau_k^1)x(\tau_k^1)$ we come to

$$-x(\tau_k^o)A(\tau_k^o)\exp(A(\tau_k^o)\tau_k^1) = A^v(\tau_k^1)x(\tau_k^o)[2-\exp(A(\tau_k^o)\tau_k^1)],$$

or to the connection of both matrixes $A^v(\tau_k^1)$ and $A(\tau_k^1)$ (at the interval end) with the matrix $A(\tau_k^o)$ (at the interval beginning) in the forms(6.6b) for the closed system:

$$A^v(\tau_k^1) = -A(\tau_k^o)\exp(A(\tau_k^o)\tau_k^1)[2-\exp(A(\tau_k^o)\tau_k^1)]^{-1}$$

and for the system (6.1):

$$A(\tau_k^1) = A(\tau_k^o)\exp(A(\tau_k^o)\tau_k^1)[2-\exp(A(\tau_k^o)\tau_k^1)]^{-1}. \quad (6.7c)$$

*Proof* (2) follows from relations (6.7a) and (6.7c), which under applying controls $v(\tau_k^1) = -2x(\tau_k^1)$ and $v(\tau_{k+1}^1) = -2x(\tau_{k+1}^1)$ change both matrixes signs according to (6.6c). *Example* A.1. illustrates this.

*Proof* (3) follows from relations

$$\dot{x} = A(x+v) = a^u, v_o = -2x(\tau), a^u(x(\tau)) = Ax(\tau) + A(-2x(\tau)) = -Ax(\tau), a^u(v_o) = Av_o/2;$$

which at $v_o(\tau+o) = -2x(\tau+o) \to 0$ lead to $a^u(v_o) \to 0$. •

Proposition 6.3. (1).The constraint Eq.

$$\frac{\partial S_c}{\partial t}(t) = b\frac{\partial X}{\partial x}(t) \qquad (6.8)$$

which at the moment $t=\tau$ of imposing the constraint takes form

$$b\frac{\partial X}{\partial x}(\tau) = -(a^u)^T X(\tau), (6.8a)$$

leads to its connection with Hamiltonian (5.8a) on extremals $H = -\frac{\partial S_o}{\partial t} = 1/2(a^u)^T(2b)^{-1}a^u$ at the same moment:

$$\frac{\partial S_c}{\partial t}(\tau) = -2H(\tau). \qquad (6.8b)$$

(2). The constraint's information contribution up to the moment $t=\tau$ equals to the Hamiltonian's contribution at the moment $t=\tau$ :

$$\Delta S_c(\tau) = -1/2\int_\tau A(t)dt = -2H(\tau)\tau, \qquad (6.8c)$$

where $A = (\lambda_i), i = 1,...n.$ (6.8d)

*Proofs.* (1). Substitution $X = (2b)^{-1}a^u$ in (6.8) from (5.2,5.3) brings

$\frac{\partial S_c}{\partial t}(\tau) = -(a^u(\tau))^T(2b(\tau))^{-1}a^u(\tau)$, which leads to (6.8b).

(2). Using Eqs $\dot{x} = 2bX, \dot{x} = br^{-1}x,$ we have $X = 1/2r^{-1}x$ and $\frac{\partial X}{\partial x} = 1/2r^{-1}$, which, at $A(t) = -b(t)r^{-1}(t)$, brings $b\frac{\partial X}{\partial x} = 1/2br^{-1} = -1/2A$ and then (6.8d).•

Corollary. 6.3. A total derivation (5.6) during a dynamic movement up to the constraint imposition leads to a *balance* Eq: $\frac{\partial S}{\partial t} = \frac{\partial S_c}{\partial t} + \frac{\partial S_o}{\partial t} = 3\frac{\partial S_o}{\partial t} = -3H$ . (6.8e)

*Proposition* 6.4.

(1)-Holding constraint (6.4) for the considered conjugated Hamilton Eqs. brings the connections of complex conjugated eigenvalues ($(\lambda_i(\tau), \lambda_j(\tau))$ of this equation's operator $A^v(\tau)$ at each moment $\tau = \tau_k^1$ to the forms

$$\operatorname{Re}\lambda_i(\tau_k^1) = \operatorname{Re}\lambda_j(\tau_k^1),\ \operatorname{Im}\lambda_i(\tau_k^1) = \operatorname{Im}\lambda_j(\tau_k^1) = 0; \qquad (6.9)$$

(2)-Joint solutions of Eqs (6.4) and (6.4a) under applying control $v(\tau_k^o) = -2x(\tau_k^o)$, starting at $\tau_k^o$, indicates the moment $\tau_k^1$ when the solution of the constraint equation approaches the process' punched locality, and control $v(\tau_k^o)$ should be turned off, transferring the extremal movement to a random process at the following moment $\tau_{k+1}$;

(3)-At the moment of turning the constraint off: $(\tau_k^1 + o) \to \tau_{k+1}$, the control joins both real eigenvalues (6.9) of the conjugated $A^v(\tau)$ unifying these two eigenvalues:

$$\operatorname{Re}\lambda_{ik}(\tau_k^1 + o) = \operatorname{Re}\lambda_i(\tau_k^1) + \operatorname{Re}\lambda_k(\tau_k^1) \qquad (6.9a)$$

and, hence, joins each $i, j$ dimensions, corresponding these conjugated eigenvalues;

(4)-Each moment $\tau_k^1$ of the control's turning off is found from the completion of the constraint Eq. (6.4) in the form (6.9) (at this moment) under the starting control $v(\tau_k^o) = -2x(\tau_k^o)$;

(5)-Within time interval $\tau_k^1 - \tau_k^o = t_k$, the dynamic movement, approximating the diffusion process with minimal entropy functional on the trajectories, moves to the locality of imposing constraint with a maximal information speed of decreasing this functional; and by reaching this locality it minimizes this speed.

At the moment $\tau_k^1$, the extremal approaches the diffusion process with a maximal probability.

*Proof* (1).Specifying constraint equations (6.4), (6.4a) at the ending moment $\tau_k^1$ of interval $t_k, k=1,...,m$, we have

$X_i = A_i(x_k+v_k)(2b_i)^{-1}, X_k = A_k(x_i+v_i)(2b_k)^{-1}, \ r_{ik} = E[(x_i+v_i)(x_k+v_k)] = r_{ki}$, (6.9b)

$\frac{\partial X_i}{\partial x_k} = (2b_i)^{-1}A_i, A_i = -(r_i)^{-1}b_i; \ \frac{\partial X_k}{\partial x_i} = (2b_k)^{-1}A_k, A_k = -(r_k)^{-1}b_k.$

Substitution these relations in the constraint's Eq.:

$$\frac{\partial X_i}{\partial x_k} = -2X_iX_k = \frac{\partial X_k}{\partial x_i} \quad (6.9c)$$

leads to $E[X_iX_k] = 1/4r_{ii}^{-1}r_{ik}r_{kk}^{-1}, r_{ii} = E[x_i^2(t)], r_{kk} = E[x_k^2(t)]$- at the central part of Eq. (6.9c), and to $E[\frac{\partial X_i}{\partial x_k}] = 1/2r_{ii}^{-1}, E[\frac{\partial X_k}{\partial x_i}] = 1/2r_{kk}^{-1}$- on the left and right sides of the above Eq.

Suppose, imposing the constraint brings the connection of conjugate variables in the form $E[X_iX_k] \neq 0$, which holds true at the moment $\tau_k^1$ of the constraint starts.

It is seen that, at any finite auto-correlations $r_{ii} \neq 0, r_{kk} \neq 0$, this connection is holding only if $r_{ik}(\tau_k^1) \neq 0$.

The vice versa: $E[X_iX_k] = 0$ leads to $r_{ik} = 0$ at any other moments when the constrain is absent.

Joining the above relations for the central and both sides of (6.9b), we get

$1/2r_{ii}^{-1} = 1/2r_{ii}^{-1}r_{ik}r_{kk}^{-1} = 1/2r_{kk}^{-1}, or \ \ r_{ii}(\tau_k^1) = r_{kk}(\tau_k^1) = r_{ik}(\tau_k^1)$.

This means that the initial auto-correlations $r_{ii}(t), r_{kk}(t)$ become the mutual correlations $r_{ik}(t)$ at the moment $t = \tau_k^1$ of the constraint imposing, and random states $\tilde{x}_i(t), \tilde{x}_k(t)$ of the dimensions $i,k$ become connected.

Taking the derivations: $\dot{r}_{ii}(\tau_k^1) = \dot{r}_{kk}(\tau_k^1)$ leads to $b_i(\tau_k^1) = b_k(\tau_k^1)$, and then by multiplying this equation on $r_{ii}(\tau_k^1)^{-1} = r_{kk}(\tau_k^1)^{-1}$ we get $\dot{r}_{ii}(\tau_k^1)r_{ii}(\tau_k^1)^{-1} = \dot{r}_{kk}(\tau_k^1)r_{kk}(\tau_k^1)^{-1}$ which, by following (6.9a), leads to

$$A_i(\tau_k^1) = A_k(\tau_k^1), \quad (6.10)$$

or for the matrix's eigenvalue $i,k$, considered at the same moment $\tau_k^1$, we get

$$\lambda_i(\tau_k^1) = \lambda_k(\tau_k^1) \ . \quad (6.10a)$$

Thus, imposing the constraint brings the connection of model's processes in the form (6.10a).

For a related complex conjugated eigenvalues, corresponding to a Hamilton Eqs. of a dynamic movement: $\lambda_i(\tau_k^1) = \alpha_i(\tau_k^1) + j\beta_i(\tau_k^1), \lambda_i^*(\tau_k^1) = \alpha_i(\tau_k^1) - j\beta_i(\tau_k^1),$

satisfying (6.10a) in the form $\lambda_i(\tau_k^1) = \lambda_i^*(\tau_k^1)$, we come to

$$\operatorname{Im}\lambda_i(\tau_k^1) = \beta_i(\tau_k^1) = \operatorname{Im}[\lambda_i^*(\tau_k^1)] = -\beta_i(\tau_k^1), 2\beta_i(\tau_k^1) = 0, \operatorname{Im}\lambda_i(\tau_k^1) = 0, \quad (6.10b)$$

and $$\operatorname{Re}\lambda_i(\tau_k^1) = \operatorname{Re}\lambda_i^*(\tau_k^1). \quad (6.10c)$$

This prove (6.9) in (1).

*Proof* (2).After applying the control at the moment $\tau_k^o$ to both (5.5a) and (6.4) and using Eqs (6.6c):

$$b(t)\frac{\partial X(t)}{\partial x(t)}+A(\tau_k^o,t)x(\tau_k^o,t)(2-\exp(A(\tau_k^o)t))^T X^T(t)+1/2a^u(\tau_k,\tau_k^o,t)(2b(t)^{-1}(a^u(\tau_k,\tau_k^o,t)))^T=-\frac{\partial S}{\partial t}(t), (6.11)$$

the model moves to fulfill (5.4) by moment $\tau_k^1$ with the above maximal entropy speed decreasing the entropy functional. When the constraint is turning on, we get

$$\frac{\partial X(\tau_k^1)}{\partial x(\tau_k^1)}+2A(\tau_k^o,\tau_k^1)x(\tau_k^o,\tau_k^1)(2-\exp A(\tau_k^o)\tau_k^1)X^T(\tau_k^1)=0\ , \qquad (6.11a)$$

and the entropy speed reaches its minimum.

Thus, completion of (6.10a) *indicates* the moment $\tau_k^1$ when the solution of the constraint equation approaches the process' punched locality, and control $v(\tau_k^o)$ should be turned off, transferring the extremal movement to a random process at the following moment $\tau_{k+1}$.

It seen that satisfaction of any (6.10),(6.10a-c) actually follows from imposing the constraint (starting at $\tau_k^o$), which by the moment $\tau_k^1$ reaches the $(\tau_{k+1}-o)$ locality. Therefore, the conditions (6.10),(6.10a-c) are indicators of the (6.11a) completion, and hence, can be used to find the moment $\tau_k^1$ of turning the control off.

*Proof* (3). As it follows from Col.6.1, both real eigenvalues (6.9) should be transformed to a real component of diffusion matrix (according to (6.3)) after turning off the constraint. This requires completion of (6.9a).

*Proof* (4)- (5). The impulse control action at the moment $\tau_k^1$ transfers the dynamic movement to the punched locality and activates the constraint in the form (5.9c) on at this locality. At this moment, dynamic movement approaches the diffusion process with minimal entropy speed and the corresponding maximal probability. As it follows from Prop. 5.2 and Col(5.2), state $x(\tau_k^1)$ holds a *dynamic* analogy of random state $x(\tau_{k+1}-o)$ with a maximal probability, which is used to form a stepwise control (6.2), starting a next extremal segment; while both stepwise controls (forming of the impulse control) connects segments between the boundary points. The extremals provide a dynamic approximation of the diffusion process and allows its modeling under a currently applied optimal control. •

<u>Proposition 6.5</u>

(1)-Math expectation of the IPF derivation in (5.5a) (taken on each extremal by the moments $\tau=\tau_k$ ) and the matrix's $A(\tau)$ eigenvalues for each of the model dimension $i=1,...,n,$ are connected via the equations

$$E[\frac{\partial\tilde{S}}{\partial t}(\tau)]=1/4Tr[A(\tau)]\ , \qquad (6.10)$$

$$A(\tau)=(\lambda_i(\tau)),\lambda_i(\tau_k)=4E[\frac{\partial\tilde{S}_i}{\partial t}(\tau_k)], i=1,...n, k=1,...m\ . \qquad (6.10a)$$

The *proof* follows from substituting to (5.6) the relations (6.1), (6.3) that leads to Eq

$$E[-\frac{\partial\tilde{S}}{\partial t}(\tau)]=1/2E[(x(\tau)+v(\tau))^T A(\tau)^T(2b)^{-1}A(\tau)(x(\tau)+v(\tau))],$$

from which at $A(\tau)=-b(\tau)r_v^{-1}(\tau)$ we get (6.10); and for each *i*-dimensional model's eigenvalue with $\tau=\tau_k$ we come to (6.10a), where for a stable process $A(\tau)<0$ •

(2)- The IPF for a total process, with $A(\tau)=-1/2\sum_{i=1}^{n}\dot{r}_i(\tau)r_i^{-1}(\tau),(r_i\ )=r$, measured at the punched localities, acquires the form

$$I^{p}_{x_t} = -1/8\int_{s}^{T} Tr[\dot{r}r^{-1}]dt = -1/8Tr[\ln(r(T)/\ln r(s)],(s=\tau_o,\tau_1,...,\tau_n=T). \quad (6.11)$$

(3)-An elementary entropy increment $S_i^{\delta}$ between the nearest segments' time interval $(\tau_k^i - o, \tau_k^i + o)$ corresponding to the cut-off (Sec3) is

$$S_i^{\delta} = -1/8\int_{\tau_k^i - o}^{\tau_k^i + o} r_i^{-1}(t)\dot{r}_i(t)dt = -1/8\ln[r_i(\tau_k^i + o)/r_i(\tau_k^i - o)]. \quad (6.11a)$$

Considering approximation of these moments by those at the end of a pervious segment: $\tau_k^i - o = \tau_k^{1i}$ and the beginning of a following segment: $\tau_k^i + o = \tau_k^{oi+1}$, we get

$$S_i^{\delta} = 1/8[\ln r_i(\tau_k^{1i}) - \ln r_i(\tau_k^{oi+1})], \quad (6.11b)$$

where the correlations are taken sequentially in the times course, from the moment $\tau_k^{1i}$ to the moment $\tau_k^{oi+1}$.

This quantity of information, evaluated in (Sec.3a): $S_i^{\delta} \cong 0.5Nats$, delivers the process' hidden information. •

***7. The model's information invariants.***

<u>*Proposition*</u> <u>7.1.</u> The constraint's information invariant $E[\Delta\tilde{S}_{ic}] = \Delta S_{ic} = inv$ (5.10) during interval $t_k$ of applying controls $v(\tau_k^o) = -2x(\tau_k^o)$ leads to the following three invariants:

$$\lambda_i(\tau_k^o)\tau_k^o = inv = i_1 \quad (7.1)$$

$$\lambda_i(\tau_k^1)\tau_k^1 = inv = i_2, \quad (7.1a)$$

$$\lambda_i(\tau_k^o)t_k = inv = i_3, \quad (7.1b)$$

and to their connections in the forms

$$i_1 = 2i_2, \quad (7.2)$$

$$i_2 = i_3 \exp i_3 (2 - \exp i_3)^{-1}, \quad (7.2a)$$

where $\lambda_i(\tau_k^o)$ and $\lambda_i(\tau_k^1)$ are complex eigenvalues of the model matrix $A = (\lambda_i)_{i=1}^n$, taken at the at the moments $\tau_k^o$ and $\tau_k^1$ of a segment's time interval $t_k$ accordingly. • Detailed proof is in Lerner [23].

The *proof* uses the constraint's invariant information contribution (6.8c) in the form :

$$\Delta S_{ic}(\tau) = -1/2\int_{\tau_k^{oi}}^{\tau_k^{1i}} \lambda_i(t)dt = -1/2(\lambda_i(\tau_k^1)\tau_k^{1i} - \lambda_i(\tau_k^{oi})\tau_k^{oi}) = inv, \quad (7.3)$$

which according to (6.8d) is connected to the Hamiltonian at this moment: $-2H_i(\tau_k^{1i})\tau_k^{1i} = 1/2\lambda_i(\tau_k^1)\tau_k^{1i}$ in the form

$$(\lambda_i(\tau_k^1)\tau_k^{1i} - \lambda_i(\tau_k^{oi})\tau_k^{oi}) = inv = -\lambda_i(\tau_k^1)\tau_k^{1i}, \quad (7.4)$$

which leads to $\lambda_i(\tau_k^1)\tau_k^{1i} = 1/2\lambda_i(\tau_k^{oi})\tau_k^{oi} = inv$. (7.4a)

This *proves* both (7.1) , (7.1a), and also (7.2).

To prove (7.1b) we use the eigenvalues' function

$$\lambda_i(\tau_k^1) = \lambda_i(\tau_k^o)\exp(\lambda_i(\tau_k^o)\tau_k^1)[(2-\exp(\lambda_i(\tau_k^o)\tau_k^1)]^{-1} \quad (7.5)$$

following from (6.6a) after applying the control $v_i(\tau_k^o) = -2x_i(\tau_k^o)$ at $A = (\lambda_i), i = 1,...,k,...,n$.

Multiplying both sides of (7.5) on $\tau_k^1$ and substituting invariant (7.4) we obtain

$\lambda_i(\tau_k^1)\tau_k^1 = \lambda_i(\tau_k^o)\tau_k^1 \exp(\lambda_i(\tau_k^o)\tau_k^1)[(2-\exp(\lambda_i(\tau_k^o)\tau_k^1)]^{-1} = inv$ . (7.6) •

Considering the real eigenvalues for the above complex eigenvalues at each of the above moments:
$\operatorname{Re}\lambda_i(\tau_k^o) = \alpha_i(\tau_k^o)$ , $\operatorname{Re}\lambda_i(\tau_k^1) = \alpha_i(\tau_k^1)$, we come to the real forms of invariant relations in (7.1-7.2):

$\alpha_i(\tau_k^1)\tau_k^1 = \mathbf{a}_i$, $\alpha_i(\tau_k^o)\tau_k^o = 2\mathbf{a}_i$, $\alpha_i(\tau_k^o)\tau_k^1 = \mathbf{a}_{io}$ , (7.7)

Applying relations (7.1), (7.1a-b) for the complex eigenvalues' imaginary parts:
$\operatorname{Im}\lambda_i(\tau_k^1) = \beta_{ik}$ , $\operatorname{Im}\lambda_i(\tau_k^o) = \beta_{iko}$ brings the related imaginary invariants:

$\beta_i(\tau_k^1)\tau_k^1 = \mathbf{b}_i$ , $\beta_i(\tau_k^o)t_k = \mathbf{b}_{io}$ . (7.8)

*Remark* in A.1.defines an imaginary invariant at turning the real eigenvalues to zero, and *Example* A2. defines an invariant for a dynamic model with only imaginary conjugated eigenvalues.

***8. The evaluations of the model's information contributions by the invariants.***

From (8.3) we have the entropy increment, expressed by the invariants

$\Delta S_i^\delta = -1/2(\mathbf{a}_i - 2\mathbf{a}_i) = 1/2\mathbf{a}_i$ . (8.1)

And following (6.8c) we get $\Delta S_{io} = 3/2\mathbf{a}_i$ .

If we evaluate $S_{io}$ using invariant $\mathbf{a}_{io} : \Delta S_{io} = \mathbf{a}_{io}/2$ , at $\mathbf{a}_i = \mathbf{a}_{io} \exp \mathbf{a}_{io}(2-\exp \mathbf{a}_{io})^{-1}$, then the ratio $\Delta S_{io} / \Delta S_i^\delta \cong 3$ should be correct. Let us check it.

We have $\Delta S_{io} / \Delta S_i^\delta \cong \mathbf{a}_{io} / \mathbf{a}_i = \exp(-\mathbf{a}_{io})(2-\exp \mathbf{a}_{io})$ . (8.2)

This relations, at $\gamma \to (0.0-0.6)$ takes the values between $3.28-2.63$, with $\exp(-\mathbf{a}_{io})(2-\exp \mathbf{a}_{io}) \cong 3$, at $\gamma \cong 0.5$, which satisfies to the above balance. Therefore *the balance is* estimated by $\mathbf{a}_{io} \cong 0.7 \cong \ln 2$ , at $\gamma \cong 0.5$ .

Invariants $\mathbf{b}_{io}$ and $\mathbf{b}_i$ evaluate the model segment's information, generated by the eigenvalues imaginary components. Eq. (6.10b,c) impose the restriction on the solution of (6.3a) during each interval $t_k$ of extremal segment, which we express through the equation for invariant $\mathbf{a}_{io}$ and the starting eigenvalue's ratio $\gamma_i = \beta_i(\tau_k^o)/\alpha_i(\tau_k^o)$ :

$2\sin(\gamma_i \mathbf{a}_{io}) + \gamma_i \cos(\gamma_i \mathbf{a}_{io}) - \gamma_i \exp(\mathbf{a}_{io}) = 0.$ (8.3)

This equation allows us to find *function* $\mathbf{a}_{io} = \mathbf{a}_{io}(\gamma_i)$ for each starting eigenvalue $\alpha_{iko} = \alpha_i(\tau_k^o)$, satisfying constraint (6.4), and also get $\gamma_i(\alpha_{iko})$, indicating the dependence of $\gamma_i$ on the identified $\alpha_{iko}$ **.**

At each fixed $\gamma_i = \gamma_{i*}$, invariant $\mathbf{a}_{io}(\gamma_{i*})$ determines quantity of information needed to develop the information dynamic process approximating initial random process with a maximal probability.

Since real eigenvalues (6.9a) are doubling-up, invariant $S_{io}$ is evaluated by $\mathbf{a}_{io}(\gamma_i)$, and other invariants are doubling accordingly. Then the quantity of external information $I_s^i$, following from (6.11b), that is needed for information dynamics, is evaluated by invariant:

$I_s^i = \mathbf{a}_i(\gamma_i)$ . (8.4)

This directly connects the required external information with information necessary for information dynamics.
Information (8.4), evaluated in Sec.3, is an equivalent of the information invariant $0.5\mathbf{a}_{oi}$ at $\mathbf{a}_{oi} \cong 1 Nats$ **.**

Let us find optimal $\gamma_{io}$ corresponding to an exteme of the information invariant (following from Eqs.(6.7c)), (7.2):

$\mathbf{a}_i(\gamma_i) = \mathbf{a}_{io}(\gamma_i) \exp \mathbf{a}_{io}(\gamma_i)(2-\exp \mathbf{a}_{io}(\gamma_i))^{-1}$ , (8.5)

which would be necessary to develop an *optimal* dynamic process approximating the measured information process.
Using a simple indications $\mathbf{a}_{io}(\gamma_i) = y(x), \mathbf{a}_i(\gamma_i) = z(y(x))$ , we get

$$etrz = dz/dy(dx) = [(2-\exp y)][(dy/dx)\exp y + (dy/dx)y\exp y[(2-\exp y)]^{-2}$$
$$-y\exp y(-(dy/dx)\exp y)[(2-\exp y)]^{-2} = 0;$$
$$(dy/dx)(1+y)\exp y + (dy/dx)y\exp 2y[(2-\exp y)]^{-1} = 0, \exp y \neq 2,$$
$$(1+y) + y\exp y[(2-\exp y)]^{-1} = 0, (dy/dx) \neq 0, y \neq 0.$$

We come to Eq. $2(1+y) = \exp y$, whose solution $y \cong -0.77 = \mathbf{a}_{io}(\gamma_{io})$ (8. 6)

allows us to find $\gamma_i$ equivalent to (8.6): $\gamma_{io} \to 0$, and using (8.6) to get $\mathbf{a}_i(\gamma_{io}) \cong 0.23$. (8.7).

This is a minimal $\mathbf{a}_i(\gamma_{io})$ in diapason of admissible $\gamma_i \to (0,1)$, while at

$\gamma_{io} \to 0, \beta_i(\tau_k^o) \to 0$, at $\mathbf{b}_{io} \to 0, \mathbf{a}_{io} \to \max$. (8.8)

The minimal $\mathbf{a}_i(\gamma_{io})$, obtained from random process, corresponds to implementation of the minimax principle during interval of applied impulse control, which enables delivering a maximum information.

Moreover, the minimal $\mathbf{a}_i(\gamma_{io})$ provides a maximum information $\mathbf{a}_{io}(\gamma_{io})$ for optimal dynamic process.

The information values of the above invariants $\mathbf{a}_{io}$ and $\mathbf{a}_i$ correspond to 1.1 bit and 0.34 bits accordingly.

(These results, following from the IPF, do not apply classical information theory).

The dynamic information $\mathbf{a}_{io}(\gamma_{io})$ includes information needed by a stepwise control starting this *i*-segment, which is evaluated by $\mathbf{a}_{ic}(\gamma_{io})$. Another stepwise control (as a first part of the impulse control), turning the constraint off, requires the same information $\mathbf{a}_{ic}(\gamma_{io})$. Since, the invariant $S_{io}$ is evaluated by invariant $\mathbf{a}_{io}(\gamma_{io}) = \alpha_i(\tau_k^o)t_k^i$, which includes $2\,\mathbf{a}_{ic}(\gamma_{io})$, a total needed dynamic information is

$S_{io}^o = \mathbf{a}_{io}(\gamma_{io}) - 2\mathbf{a}_{ic}(\gamma_{io})$. (8.9)

This information should be extracted from a random process, which is evaluated by $I_{so}^i = \mathbf{a}_i(\gamma_{io})$ .

We get balance information at $S_{io}^o = \mathbf{a}_i(\gamma_{io}) = \mathbf{a}_{io}(\gamma_{io}) - 2\mathbf{a}_{ic}(\gamma_{io})$, (8.9a)

from which get information needed for a single stepwise control:

$\mathbf{a}_{ic}(\gamma_{io}) = 1/2(\mathbf{a}_{io}(\gamma_{io}) - \mathbf{a}_i(\gamma_{io}))$, (8.9b)

evaluated for the optimal process by $\mathbf{a}_{ic}(\gamma_{io}) \cong 0.27Nat$, with the total impulse control information

$2\mathbf{a}_{ic}(\gamma_{io}) \cong 0.54Nat$, (8.9c)

which, we assume, provides an extractror (observer), which also generates the information dynamics, which, according to $S_{io}^o$ , is evaluated by the invariants (8.8a). This would bring the total information

$S_{io} = \mathbf{a}_{io}(\gamma_{io}) = 0.77bit \neq 3\mathbf{a}_i(\gamma_{io})$, while the equality in this balance relation is reached at $\gamma_i \to 0.5$.

At unknown control information, it can be evaluated by a dynamic invariant in the form

$2\mathbf{a}_{ic}(\gamma_i) \cong \mathbf{a}_{io}^2(\gamma_i)$ . (8.10)

For the optimal dynamics we get $\mathbf{a}_{io}^2(\gamma_{io}) \cong 0.593Nat$ which is closed to (8.9c). In average, we have

$\mathbf{a}_{ic}(\gamma_i) \approx \mathbf{a}_i(\gamma_i) \approx 1/3\mathbf{a}_{io}(\gamma_i)$ . (8.10a)

### *9. The multi-dimensional information dynamics satisfying the VP.*

Let us have a $n$-dimensional spectrum of the model operator with complex conjugated eigenfunctions $A_t = A(\lambda_{it}), i = 1,...,n$, starting simultaneously at the moment $t = t_o$ under applying a stepwise control $v = v(t_o)$, with initial non equal eigenvalues at this moment $A_o(\lambda_{io}), i = 1,...,n$ ) – corresponding two conjugated macrodynamic processes starting at a beginning of each extremal segment .

During a dynamic movement at each segment, satisfying the VP minimax principle, its initial complex eigenvalue $\{\lambda_{io}\}, i=1,...,n$ is transformed in a real eigenvalue $\{\alpha_{it}\}, i=1,...,n$ by the end of the segment's time interval $t_k^i$, according to the invariant relation (8.5): $\alpha_{it}=\alpha_{io}\exp\mathbf{a}_{io}(2-\exp\mathbf{a}_{io})^{-1}$ , (9.1) where $\{\alpha_{io}\}, i=1,...,n$ are real components of $\{\lambda_{io}\}$ that carry the corresponding frequency of the spectrum, and $\mathbf{a}_{io}=\alpha_{io}t_k$ is the model's invariant. Applying these optimal dynamics, we will prove the following Proposition.

*Proposition 9.1.* Current time course of the controllable information process, satisfying the VP, is *accompanied* with a sequential *ordering of both macromodel's information spectrum* and *time intervals* of the extremal's segments.

Specifically, for the ranged spectrum of the model's real parts of eigenvalues: $A_o(\alpha_{1o},\alpha_{2o},\alpha_{3o},..,\alpha_{io},...,\alpha_{no})$, selected at a beginning of each process' *n* segments, where $\alpha_{1o}$ holds a maximal information frequency and $\alpha_{no}$ holds a minimal information frequency, it is *required to prove*:

(1) That, at a given invariant $\mathbf{a}_{io}$, the $\alpha_{1o}$ is selected from the process' *shortest* segment's time interval, and $\alpha_{no}$ is selected from the process' *longest* segment's time interval; and

(2) The process' current time course consists of a sequence of the segment's ordered time intervals $t_k^1, t_k^2, t_k^3,...,t_k^i,...,t_k^n$, with $t_k^1$ as a shortest segment's time interval and $t_k^n$ as a longest segment's time interval, while $t_k^1$ is the *first* time interval at the process beginning.

*Proof* (1) follows from invariant $\mathbf{a}_{io}=\alpha_{io}t_k^i$, $t_k^i=\tau_k^i-\tau_{k-1}^i$, which implies that each maximal $\alpha_{io}(\tau_{k-1}^i)$ (with a fixed $\tau_{k-1}^i$) corresponds to a minimal $t_k^1$, or vice versa, each minimal $\alpha_{no}(\tau_k^n)$ is selected from a maximal $t_k^n$.

P*roof* (2) is a result of reaching each local minima of the VP functional's derivation at the moment $\tau_k^i$ of each segment's end:

$$E\,|\frac{\partial S_{ik}}{\partial t}(\tau_k^i)|=|\alpha_{ik}(\tau_k^i)|\rightarrow \min, i=1,...,n, \qquad (9.2)$$

(along the time course of the optimal process with $(\tau_k^i,\tau_{k+1}^{i+1},...\tau_m^n), i=1,...n, k=1,...,m$)) and reaching a global minima for this functional's derivations at the process end: $E\,|\frac{\partial S_{kn}}{\partial t}|=|\sum_{i=1}^{n}\alpha_{ik}|\rightarrow \min$. (9.2a)

Keeping a minimum of derivation (9.2), according to (9.1), preserves this minimum along each segment for each fixed $\alpha_{io}$. With a minimal path functional on each fixed time interval $t_k^i=\tau_k^i-\tau_{k-1}^i$, this brings the minimal increments of this functional and adds it to each following segment's minima according to (9.2), allowing to reaching (9.2a) by the end. Each current minimal eigenvalue, selected by the VP minimax principle, has a maximum among other minimal eigenvalues, which would be selected from this spectrum. From this it follows a sequential declining of these minimal increments along the time course of the optimal process at the moments $(\tau_k^i,\tau_{k+1}^{i+1},...\tau_m^n), i=1,...n, k=1,...,m$:

$$\min\alpha_{i+1t}(\tau_{k+1}^{i+1})<\min\alpha_{it}(\tau_k^i)\ \ or\min\alpha_{it}(\tau_k^i)>\min\alpha_{i+1t}(\tau_{k+1}^{i+1}),\ \tau_{k+1}^{i+1}>\tau_k^i,.... \qquad (9.3)$$

In such a sequence, each minimal eigenvalue $\alpha_{it}(\tau_k^i)$ holds a maximum with regard to all following minimal eigenvalues. From (9.1, 9.2), it follows that minimum of $\alpha_{it}$ leads to minimum for $\alpha_{io}$ (at a fixed invariant).

Since each sequence of $\alpha_{it}(\tau_k^i)$ corresponds to related sequence of minimals $\alpha_{io}(\tau_{k-1}^i)$ along the time course of the optimal process $\alpha_{io}(\tau_k^i,\tau_{k+1}^{i+1},...\tau_m^n)$, both $\alpha_{it}(\tau_k^i)$ and $\alpha_{io}(\tau_{k-1}^i)$ will be orderly arranged during this time course. Then, this ordered connection holds true for all $A_t(\alpha_{1k},\alpha_{2k},\alpha_{3k},..,\alpha_{ik},...,\alpha_{nk})$ and related $A_o(\alpha_{1o},\alpha_{2o},\alpha_{3o},..,\alpha_{io},...,\alpha_{no})$

during the optimal process' time course. Sequential ordering of both eigenvalues $\alpha_{it}(\tau_k^i)$ and $\alpha_{io}(\tau_{k-1}^i)$ leads to the ordering of the corresponding segment's time intervals $\{t_k^i\}$, starting with its minimal $t_k^1$, at the process beginning with its maximal $\alpha_{1o}$. The proposition is proved, confirming also the initial assumption of the ranged spectrum.•

*Therefore,* ordering of the initial eigenvalues satisfies to the minimax principle, which selects sequentially such of the following eigenvalue that brings a maximal eigenvalue among all other minimal eigenvalues to this optimal spectrum.

The informational dynamics in time-space, described by a sequence of the IPF space distributed extremal segments (Lerner [23,26]), form spiral trajectories, located on a conic surface, while each segment represents a three-dimensional extremal. The segment's conjugated dynamic trajectories form an opposite directional double spirals **(**Fig. 2) rotating on the surfaces of the same cone. On the cone's vertex, the opposite trajectories join together with their equal real eigenvalues. Such an optimal spiral geometry follows from each pair of the distributed in space the model's Hamilton Eqs, whose conjugated space solutions-waves decrease exponentially, approaching to their merge at the cone's vertex.

The implementation of the IPF minimax principle leads to a sequential assembling of a manifold of the process' extremals (Fig.1b) in elementary binary units (doublets) and then in triplets, producing a spectrum of coherent frequencies. The manifold of the extremal segments, cooperating in the triplet's optimal assemble, are shaped by a sequence of the spirals conic structures, whose cones' vertexes form the nodes of an information network (IN) with a hierarchy of the IN nodes (Fig.2).

***10. Measuring and evaluation of the extracted information***

Information, extracted during a time interval $\delta\tau_k^i$, is determined by the correlations (6.11a), measured at some initial moments $\tau_k^i = \tau_o^i, \tau_k^i + \delta\tau_k^i = \tau_o^i + \delta\tau_o^i$, in the form

$$I_s = -1/8Tr[\ln(r_{ij}(\tau_o^i, \tau_o^i + \delta\tau_o^i) / r_{ij}(\tau_o^i))], i, j = 1,....,n \ , \ I_s = \sum_{i=1}^{n} I_s^i \ . \qquad (10.1)$$

Following Levi [29] we estimate the ratio

$$\tau_o^i / [(\tau_o^i + \delta\tau_o^i) - \tau_o^i)] = r_i^{1/2}(\tau_o^i) / r_i^{1/2}(\tau_o^i, \tau_o^i + \delta\tau_o^i) \qquad (10.2)$$

by $\delta\tau_o^i / \tau_o^i = [r_i\ (\tau_o^i, \tau_o^i + \delta\tau_o^i) / r_i\ (\tau_o^i)]^{1/2}$ . (10.2a)

The estimation holds true specifically for random state $\tilde{x}_i(\tau_o^i), \tilde{x}_i(\tau_o^i + \delta\tau_o^i)$ between the alleged moments $(\tau_o^i, \tau_o^i + \delta\tau_o^i)$ of the process's cut-off, and for the correlation $r_i\ (\tau_o^i, \tau_o^i + \delta\tau_o^i)$ (that supposable would be cut-off between these moments). Information $I_s^i$ , obtained during time interval $\delta\tau_o^i$ , is evaluated by the invariant:

$$I_s^i = \alpha_i\ (\tau_o^i)\delta\tau_o^i = \mathbf{a}_i(\gamma_i)\ , \qquad (10.\ 3)$$

Joining (10.1) ,(10.2a) and (10.3) and using invariant the invariant $\mathbf{a}_i$ for evaluations of the correlations, we have

$$[r_i\ (\tau_o^i, \tau_o^i + \delta\tau_o^i) / r_i\ (\tau_o^i)] = \exp(-8\mathbf{a}_i)\ . \qquad (10.4)$$

And then $\delta\tau_o^i / \tau_o^i = \exp(-4\mathbf{a}_i)$ , (10.4a)

which determine a relative interval of applying impulse control.

A stepwise control, starting the following extremal segment at the moment $\tau_1^{io} = \delta\tau_o^i + o$ , converts eigenvalue $\alpha_i\ (\tau_o^i)$ (in 10.3) to $\alpha_i\ (\tau_1^{io}) = |\alpha_i\ (\tau_o^i)\ |$. Then (10.4) holds form

$$I_s^i = \alpha_i\ (\tau_1^o)t_1^i / (\delta\tau_o^i / t_1^i) = \mathbf{a}_{io}\delta\tau_o^{*i},\ \delta\tau_o^{*i} = \delta\tau_o^i / t_1^i\ . \qquad (10.5)$$

This means, that getting a constant quantity information $\mathbf{a}_{io}(\gamma_{io})$, needed for optimal dynamics, requires to apply *variable* time intervals $\delta\tau_k^{*i}$ of extracting such information $I_s^i$, which would depend of the correlations that are cutting off from the random process. The moment $\tau_o^i$ of applying stepwise control we evaluate by a stepwise function

$\tau_o^i(t)=\begin{cases}0, t<\tau_o^i\\(1, t=\tau_o^i\end{cases}$ from which it follows $\delta\tau_o^i/\tau_o^i=\delta_{oi}^*(\tau_o^i=1)=\delta\tau_o^i$. (10.6)

For a minimal $I_{so}^i=\mathbf{a}_i(\gamma_{io})$, optimal time interval is $\delta\tau_o^{*i}=\mathbf{a}_i(\gamma_{io})/\mathbf{a}_{io}(\gamma_{io})\cong 0.3$, which at $\delta\tau_o^i\cong 0.4\sec$ brings the evaluation of related segment's time interval : $t_1^i=0.4/0.3\cong 1.33\sec$.

At the known invariants, we get the ratio of the eigenvalues at the end and the beginning of any segment's time interval:

At $\alpha_i(\tau_k^1)/\alpha_i(\tau_k^o)\cong\mathbf{a}_i(\gamma_{io})/\mathbf{a}_{io}(\gamma_{io}),\mathbf{a}_i(\gamma_{io})/\mathbf{a}_{io}(\gamma_{io})\cong 0.3$, (10.7)

which allows us to find both $\alpha_i(\tau_k^1)\cong 0.3\alpha_i(\tau_k^o)$ and time interval $t_k=|\mathbf{a}_{io}(\gamma_{io})|/\alpha_i(\tau_k^o)|$ for any initial real $\alpha_i(\tau_k^o)$ identified at beginning of this interval. For the considered optimal $t_1^i$, we may predict $\alpha_i^o(\tau_k^o)=|\mathbf{a}_{io}(\gamma_{io})|/t_1^i\cong 0.579$. (10.8)

For any $\gamma_i\neq\gamma_{io}$, invariant $\mathbf{a}_i(\gamma_i)$ is found from a known external information $I_s^i$ according to $\mathbf{a}_i(\gamma_i)=I_s^i$ .

Using (10.3), we find the related invariant $\mathbf{a}_{io}(\gamma_i)$ and then $\gamma_i$ based on (8.5). After that, following (8.6b) is found the needed control information $\mathbf{a}_{ic}(\gamma_i)$ .Thus, all information invariants follows from the measured external information.

## *11.The procedure of dynamic modeling and a prediction of the diffusion process*

We assume that the observed random process is modeled by Markov diffusion process with the considered stochastic equation, and the EF-IPF relations hold true.

The procedure starts with capturing from the diffussion process its correlation

$$r=E[\tilde{x}_t(t)\tilde{x}_t^T(t+\delta t)],\ r=(r_{ij}), i, j=1,..n \qquad (11.1)$$

during fixed time intervals $t=\tau,\delta t=\delta\tau,\tau=(\tau_k^i), k=0,1,2,...$, beginning with the process' initial moments $\tau_o=(\tau_o^i)$. According to (10.1-10.3) these correlations measure a quantity of information $I_s^i$ obtained on each $\delta\tau_o^i$ for any $i$ -the dimension and its invariant $\mathbf{a}_i=I_s^i$. Applying (10.1-10.3) consists of computing $r_i\ (\tau_o^i,\tau_o^i+\delta\tau_o^i)$ from (11.1) at fixed $r_i\ (\tau_o^i)$ and $\tau_o^i$ with sequential increasing increments of $\tau_o^i+\delta t$ until $\delta t=\delta\tau_o^i$ will satisfy both (10.3) and (10.4). This connects the evaluation of relative interval $\delta\tau_o^i/\tau_o^i=\delta_{oi}^*$ with information $\mathbf{a}_i$ in the form (10.4a).

Using (10.1-10.2) and (10.4a), we get jointly $\delta\tau_o^i$ ,$\mathbf{a}_i$ and then $|\alpha_i\ (\tau_o^i)|$ from (10.3). At the moment $(\tau_o^i+\delta\tau_o^i)$, a stepwise control (being a right side of impulse control) transfers the eigenvalue $\alpha_i\ (\tau_o^i)$ (computed according to (10.3)) to a dynamic model. This indicates the extraction information (10.3) from a random process during interval $\delta_{oi}^*$, where, at a beginning of this interval $\tau_o^i$, the stepwise control (as a left-side of impulse control) supposedly had been applied. Both stepwise controls provides the considered cut-off, during a *finite* time interval $\delta_{oi}^*$.

The segment's initial nonrandom states $x(\tau_o)=\{x_i\ (\tau_o^i)\}$ can be find using the correlation function $r_i(\tau_o^i)=E[x_i^2(\tau_o^i)]=x_i^2(\tau_o^i)$ at the moment $\tau_o$ of the beginning of time interval $\delta\tau_o=(\tau_o,\tau_o+\delta\tau_o)$ in the form

$$x_i(\tau_o^i)=\pm|r_i(\tau_o^i)|^{1/2}. \qquad (11.2)$$

It allows us to apply the initial optimal control

$v(\tau_o) = -2x(\tau_o), v(\tau_o) = \{v_i(\tau_o^i)\}$ (11.3)

to all primary extremal segments, starting the optimal dynamics simultaneously at the moment $\tau_o = (\tau_o^i)$ on these segments. Then, using $I_s^i$ and (10.3), the next matrix's elements is identified, and so on (A.3. Example 3).

Since correlations $r_i\ (\tau_o^i, \tau_o^i + \delta\tau_o^i)$ for different process' dimensions would be diverse , as well as $I_s^i$, the invariants $\mathbf{a}_i$, intervals $\delta\tau_o^i$, and eigenvalues $\alpha_i^k(\tau_o^k)$ will also be dissimilar.

That's why the different primary time intervals $t_i^1(\tau_o^i),\ i = 1,...,n$ of the information dynamics on each segment will form a sequence, being automatically ordered in the time course:

$t_1^1(\tau_o^1), t_2^1(\tau_o^2), t_3^1(\tau_o^3),...t_i^1(\tau_o^i),..., t_n^1(\tau_o^n)$ . (11.4)

This allows us to arrange the related $\tau_o = (\tau_o^i)$ and $\alpha_i\ (\tau_o^i)$ in the ordered series.

For example, the ordered arrangement would bring a hierarchical decrease (or increase) of these eigenvalues: $\left|\alpha_m(\tau_o^m)\right|, m = 1,..,n$ , where each successive *m*-th eigenvalue of such series is a priory unknown.

After getting the moments $t_i^1(\tau_o^i)$ , for which the sequence (11.4) will determine the moments $\tau_1 = (\tau_1^i)$ of measuring correlation (11.1), the procedure is repeating, and a next sequence of $t_i^2(\tau_1^i)$ will be obtained :

$t_1^2(\tau_1^1), t_2^2(\tau_1^2), t_3^2(\tau_1^3),..., t_i^2(\tau_1^i),..., t_n^2(\tau_1^n)$ . (11.5)

And so on, with receiving $t_1^k(\tau_k^1), t_2^k(\tau_k^2), t_3^k(\tau_k^3),..., t_i^k(\tau_k^i),..., t_n^k(\tau_k^n)$ , where each previous $t_i^{k-1}(\tau_{k-1}^i)$ will determine the following $t_i^k(\tau_k^i)$ . The ranged series of eigenvalues at the end of each $\alpha_i\ (\tau_i^k)$ are needed for sequential joining them in hierarchical information network (IN), modeling information structure of whole *n*-dimensional dynamic system(Sec.8).

Therefore, the summarized procedure consists of :

1. Simultaneous start whole system at moment $\tau_o = (\tau_o^i)$ by the applying impulse controls with computation of the process' correlations during their time intervals for each dimension;
2. Finding the quantities of information, selected at these time intervals, the information invariant, and the related eigenvalues of the potential information dynamics on each extremal segment;
3. Finding the initial dynamic states (at the end of interval $\delta\tau_k^i$) and starting the stepwise controls at these moments, which, with the known eigenvalues, initiate dynamic process at beginning of each segment.
4. Finding the time intervals of these dynamics on each segment during the random process' time course and arranging the segment's ending eigenvalues in the terms of their decreasing;
5. Turning off the stepwise control (at the segment's end) and starting measuring the process' correlations on each moment $\tau_1 = (\tau_1^i)$ determined by the end of each $t_i^1(\tau_o^i)$ ;
6. Finding new information invariants, the related eigenvalues, initial dynamic states, and turning *on* a new stepwise control on the next segments;
7. Turning o*ff* the previous stepwise controls (after finding $t_i^2(\tau_1^i)$ and begin measuring the process's correlations, find the information invariants, eigenvalues, initial dynamic states, and then start the following controls, activating information dynamics on each segment.

The procedure allows us to approximate each segment of the diffusion process by each segment of information dynamics with a maximal probability on each segment's trajectory, thereafter transforming the random process to the probability's equivalent dynamic process during a real time course of both processes.

Periodic measuring the random process' correlations allows us to renovate each segment's dynamics by the current information, extracted from the random process during its time course, at the predictable moments when dynamics coincides with the stochastics with maximal probability, closed to 1. In such approximation, each random segment's quantity information is measured by equivalent quantity on dynamic information, expressed by invariant $\mathbf{a}_{oi}$.

This invariant depends on the real eigenvalues, measured by the correlations at the moments, when these quantities are equalized, and, therefore, both segments' hold the information equivalence above.

Using the estimation of each subsequent correlations by the correlations at the end of each previous segment $r_i(t \rightarrow \tau_k^{1i})$ (which follows from preservation of a diffusion component of the random process at the considered cut-off), we can *predict* each next segment's dynamics through the eigenvalues $\alpha_i(\tau_{k+1}^i) \approx 1/2r_i^{-1}(\tau_k^{1i})\dot{r}_i(\tau_k^{1i})$ computed at the segment's ending moment $\tau_k^{1i}$, which estimates the eigenvalues at the moment $\tau_{k+1}^i$.

*Specific* of the considered optimal process *consists of the computation of each following time interval* (where the identification of the object's operator will take place and the next optimal control is applied) *during the optimal movement under the current optimal control, formed by a simple function of dynamic states.*

In this optimal *dual* strategy, the IPF optimum predicts each extremal's segments movement not only in terms of a total functional path goal, but also by setting at each following segment the renovated values of this functional's controllable shift and diffusion, identified during the optimal movement, which currently correct this goal.

The automatic control system, applying a time delay's $t_k$ (Sec.5) dynamic feedback, which is dependable on an object's current information, has been designed, patented, and implemented in practice [22,22a].

### *12. Connection to Shannon's information theory. Encoding the initial information process.*

Considering a set of discrete states $\tilde{x}(\tau^o) = \{(\tilde{x}_i(\tau_k^o)\}, i = 1,...,n; k = 1,...,m$ at each fixed moments $\tau_k^o$ along *n*-dimensional random process, and using definition of the entropy functional (2.6), we get the conditional entropy *function* for the conditional probabilities (corresponding to (1.1)) at *all* moments $\tau_k^o$ at each process dimension $S^i_{\tau_k^o}$ *and* for the whole process $S_{\tau_k^o}$ accordingly:

$$S^i_{\tau_k^o} = -\sum_{k=1}^{m} p_k[\tilde{x}_i(\tau_k^o)] \ln p_k[\tilde{x}_i(\tau_k^o)], S_{\tau_k^o} = \sum_{i=1}^{n} S^i_{\tau_k^o}, \qquad (12.1)$$

which *coincides with the Shannon entropy* for each probability distribution $p_k[\tilde{x}_i(\tau_k^o)]$, measured at each fixed $\tilde{x}_i(\tau_k^o)$.

Function (12.1) holds all characteristics of Shannon's entropy, following from the initial Markov process and its additive functional for these *states*.

For the comparison, the controllable IPF entropy, measured at the related DP's punched localities

$\tilde{x}(\tau) = \{\tilde{x}(\tau_k)\}, k = 1,...,m$

$$\tilde{S}^i_{\tau m} = \sum_{k=1}^{m} \Delta S_k[\tilde{x}(\tau_k)], \qquad (12.2)$$

(where $\Delta S_k[\tilde{x}(\tau_k)]$ is the entropy at each $\tau_k, k = 1,...m$) ;it is *distinctive* from the entropy $S^i_{\tau_k^o}$ (12.1), because:

(1)-The IPF entropy $\Delta S_k[\tilde{x}(\tau_k)]$ holds the macrostates' ( $x_i(\tau_k - o), x_i(\tau_k), x_i(\tau_k + o)$ ) *connection* through the punched locality (performed by applying the two step-wise controls), while the EF binds all random states, including the punched localities;

(2)-The distribution $p_k = p_k[\tilde{x}(\tau_k)]$ is selected by variation conditions (5.1, 5.1a) (applied to (3.7a)), as an extremal probability distribution, where a macrostate $x(\tau_k)$ estimates a random state $\tilde{x}(\tau_k)$ with a maximum probability $p_k = p_k[\tilde{x}(\tau_k)]$.

At this "sharp" probability maximum, the information entropy (3.7) reaches its local maximum at $\tau_k$ -locality, which, for the variation problem (5.1), is associated with turning the constraint (5.10) off by the controls, switching the model to the random process.
Selecting these states (with an aid of the dynamic model's control) allows an *optimal discrete filtration* of random process at all $\tau_k, k=1,...m$, where the macromodel is identified and external information is delivered.

Thus, $x(\tau_k)$ emerges as the *most informativ*e state's evaluation of the random process at the $\tau_k$ -locality.

Even though the model identifies a sequence of the most probable states, representing the most probable trajectory of the diffusion process, the IPF minimizes the EF entropy functional, defined on a *whole* diffusion process.
Therefore, the IPF implements the optimal process' functional information measure.

Each of the entropy $\Delta S_k[x_i(\tau_k - o)]$ measures the process segment's undivided information, and the entropy $\Delta S_k[x_i(\tau_k)]$ delivers *additional* information, compared to the traditional Shannon entropies, which measure the process' sates at the related discrete moments.
Here we evaluate information contribution for each segment by the segments' $\tau_k$ -locality $\Delta S_k[x_i(\tau_k)]$.

Applying the invariant's information measure to this total contribution $3\mathbf{a}_i(\gamma_i) \cong \mathbf{a}_{io}(\gamma_i)$ (which includes all information delivered with the impulse control), we get the total process $\tilde{S}^i_{\tau m}$ *estimation* by the sum of the invariants, which count both the inner segment's and control inter-segment's information:

$$\tilde{S}^i_{\tau m} \cong \sum_{k=1}^{m} \mathbf{a}_{ok}(\gamma_k), \tilde{S}_\tau = \sum_{i=1}^{n} \tilde{S}^i_{\tau m}, \qquad (12.3)$$

where $m$ is the number of the segments, $n$ is the model dimension (assuming each segment has a single $\tau_k$ -locality).

Therefore, to *predict* each $\tau_k$ - locality, where $\Delta S_k[\tilde{x}(\tau_k)]$ should be measured, we need only each invariant $\mathbf{a}_{ok}$ which estimates the IPF entropy with a maximal process' probability. Knowing *this* entropy allows encoding the *random process* using the Shannon formula for an average optimal code-word length:

$$l_c \geq \tilde{S}^o_\tau / \ln D, \qquad (12.4)$$

where $D$ is the number of letters of the code's alphabet, which encode**s** $\tilde{S}^o_\tau$ (12.4).
An elementary code-word to encode the process' segment is

$$l_{cs} \geq \mathbf{a}_{ok}(\gamma_k)\,[\text{bit}]/\log_2 D_o, \qquad (12.5)$$

where $D_o$ is a segment's code alphabet, which implements the macrostate connections.

At $\mathbf{a}_{ok}(\gamma_k \to 0) \cong 1bit,\ D_o$ =2, we get $l_{cs} \geq 1$ , or a bit per the encoding letter.

Therefore, invariant $\mathbf{a}_{ok}$ allows us both to encode the *process* using (12.4), ( 12.5) including both the segments' and the between segments' information contributions *without* counting each related entropy (12.2), and to compress each segment's random information to $\mathbf{a}_{ok}$ bits. With such optimal invariant, the quantity of encoding information for each segment will be constant, while a width of the encoding impulse $\delta\tau^i_k$ (10.4a) would depends on the quantity of external information. Since interval between these impulses $t^i_k$ , measured by invariant $\mathbf{a}_{ok}$ , will be also constant, such encoding is described by a varied pulse-width modulation**.** At any fixed $\mathbf{a}_k \neq \mathbf{a}_{ok}$ , both interval $t^i_k$ and pulse-width will be different, and the encoding is described by a pulse-amplitude modulation with a varied time. At a fixed external information the various $\delta\tau^i_k$ will generate different $t^i_k$ , which is described by pulse-time modulation.

The assigned code also encodes and evaluates both constraint and controls' actions.

Under the constraint, *each stochastic equation* (2.1) with specific functions of the drift-vector and diffusion components *encloses a potential number of the considered discrete intervals and their sequential logics.*

The selection of both the discrete intervals and their numbers is *not arbitrary* (as it is in the known methods [34, 36, others]), since each specific selection, following from the VP, is applied to the *whole* process.

The developed procedure (Secs. 10-11) [23,26,28] (see also A.3) allows building the *process hierarchical information network* (IN) (Fig.2), which, along with encoding a sequence of the process extremals in a *real* time, also ranges the segments by the values of their local entropies–invariants. The IN building includes the eigenvector's ranging and their aggregation in the elementary cooperative units (triplets), where the IN accumulated information is conserved in the invariant form. Each current IN's triplet, evaluated by the invariant quantity of information, can be encoded in its triplet's code. Thus, each process (2.1) with its space distributed hierarchical (IN) structure, which is built during the process' real time, can be encoded by the IN's individual information code.

The cooperative IN's specific consists of revealing the process *interactive dynamics*, which emerge by the dynamic binding of the process' information and the segments' macrodynamics into a *dynamic information system with the system's code hierarchy*.

Information, collected by the Entropy Path Functional (EF), is transformed to the VP selected portions of Information Path Functional (IPF), which are separated by the windows between them.

The windows give access of new information flow that can renovate each subsequent the IPF portion.

The IPF provides optimal intervals for both measuring incoming information and its consequent accumulation.

This includes local maximums of the incoming information, which are accumulated by the IPF portions, while their connection in a chain minimizes the total information. The implementation of the VP minimum for the EF leads to assembling of the above chain's portions in the information dynamic network (IN) (Fig.2). The IN hierarchical tree is formed by the sequential consolidation of each three IPF portions into the IN triplet's nodes, which are sequentially enclosed up to formation of a final IN node, which accumulates all IN enclosed information. The EF-IPF approach converts a random process' *uncertainty* into the information dynamic process' *certainty*, with its code and the IN.

The above IPF-IN information formations implement the principle of minimum for the potential information paths, connecting some initial state with subsequent formed optimal states during the information process' time course.

This minimum principle is a particular information form of the *fundamental minimum principle* in Physics [13].

*Example of the IN* information hierarchy, enclosed in the IN nodes, computed by developed software, is shown on Fig.2.

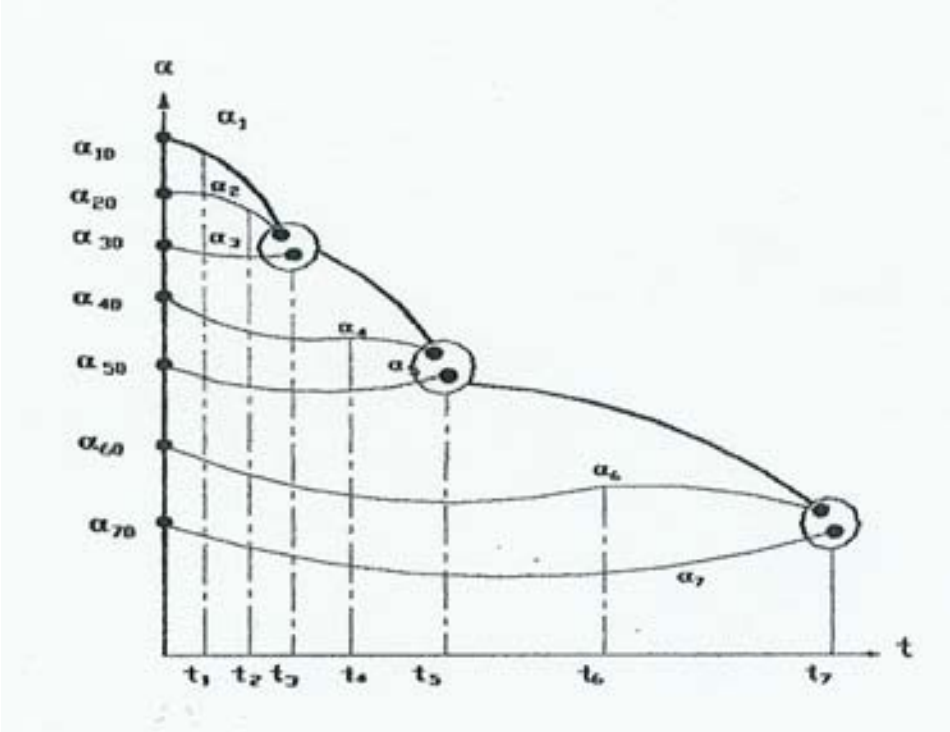

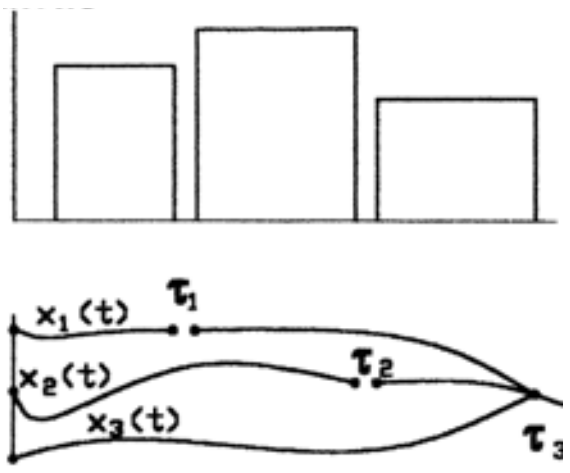

*Fig.1a.* Selection of the process' information portions (segments) and *assembling* them in a triple, *during* the initial flow's time-space movement.

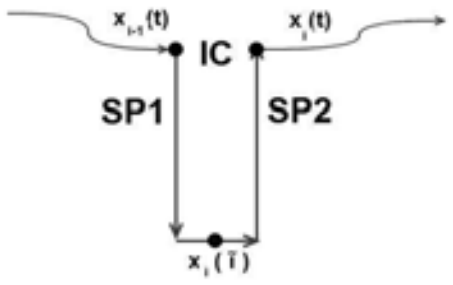

*Fig.1b.*Applying controls: IC (SP1, SP2) on the windows.

**Attachments**

**A0**. ***Controllable diffusion process and the process' information measure.***

Notion of information process is based on the definition of information for a random process, modeled by the controllable diffusion process.
Considering information as a substance being different from energy and matter, we need to find its origin and a measure, satisfying some definition and/or a law.
On intuitive level, information is associated with diverse forms of *changes* (transformations) in material and/or non-material objects of observation, expressed *universally and independently* on the changes' origin. Such universal changes are brought potentially by *random* events and processes as a set, identified via the related probability in a *probability space*, which are studied in the theory of probability, founded as a *logical* science [19]. This randomness with their probabilities we consider as a *source* of information, means that some of them, but not all of such randomness produce information. The source specifics depend on the *probability space* of a particular random event.
How can we describe a change formally and evaluate information generated by the change?
We consider generally a random process (as a continuous or discrete function $x(\omega,s)$ of random variable $\omega$ and time $s$), described by the *changes* of its *elementary* probabilities from one distribution (a priory) $P_x^a(dx)$ to another distribution (a posteriori) $P_x^p(dx)$ in the form of their transformation

$$p(x)=\frac{P_x^a(dx)}{P_x^p(dx)}, \qquad \text{(A0.1)}$$

Such a probabilistic description *generalizes* different forms of specific functional relations, represented by a sequence of different transformations, which might be extracted from the source, using the probabilities ratios (A0.1).
It is convenient to measure the probability ratio in the form of natural logarithms: $\ln[p(\omega)]$, where the logarithm of each probability $\leq 1$would be negative, or zero. That is why the considered function has the form

$$-\ln p(x)=-\ln P_x^a(dx)-(-\ln P_x^p(dx))=s_a-s_p=\Delta s_{ap}, \qquad \text{(A0.1a)}$$

represented by a difference of a priori $s_a>0$ and posteriori $s_p>0$ *entropies, which* measures *uncertainty* resulting from the transformation of probabilities for the source events, satisfying the entropy's additivity.
A *change* brings certainty or *information* if its uncertainty $\Delta s_{ap}$ is removed by some equivalent entity call information $\Delta i_{ap}:\Delta s_{ap}-\Delta i_{ap}=0$. Thus, information is delivered if $\Delta s_{ap}=\Delta i_{ap}>0$, which requires $s_p<s_a$ and a positive logarithmic measure with $0<p(x)<1$. Condition of zero information: $\Delta i_{ap}=0$ is resulted by a redundant change, transforming a priori probability to the equal a posteriori probability, or this transformation is identical–informationally undistinguished. The same way, each of the above entropies can be turned into related information. The removal of uncertainty $s_a$ by $i_a$: $s_a-i_a=0$ brings an equivalent certainty or *information* $i_a$ about entropy $s_a$. In addition, a *posteriori uncertainty* decreases (or might removes) a priory uncertainty. If $s_p = s_a$ then $s_p$ brings information $i_p = i_a$ regarding $s_a$ by removing this uncertainty, but such a change does not deliver information. At $s_p < s_a$, then $s_p$ still brings some information about $s_a$, but holds a non zero uncertainty $s_a - s_p = \Delta s_{ap} \geq 0$, which can be removed by the considered equivalent certainty - information $\Delta i_{ap}$. Uncertainty $s_a$ could also be a result of some a priori transformation, as well as uncertainty $s_p$ is resulted from a posteriori transformation.

The logarithmic measure (A0.1a) also fits to the ratio (A0.1) of Markov diffusion process' probabilities having exponential form [27], which approximates the probability's ratios for many other random processes. Because each above probability and entropy is random, to find an average tendency of functional relations (A0.1,1a) it is logically to apply a mathematical expectation:

$$E_x\{-\ln[p(x)]\} = E_{s,x}[\Delta s_{ap}] = \Delta S_{ap} = I_{ap} \neq 0, \qquad \text{(A0.2)}$$

which we call the mean *information of a random source*, being averaged by the source *events* (a probability state), or by the source *processes*, depending on what it is considered: a process, or an event, since both (A0.2) and (1.2) include also Shannon's formula for information of a states (events).

Therefore, *information* constitutes a *universal nature and a general measure of a transformation*, which conveys the changes that decrease uncertainty.

Having a *source* of information (data, symbols, relations, links, etc.) does not mean we get *information* and might evaluate its quantity and or a quality. A source only provides changes.

To obtain information, a subset of probability space, selected by formula (A0.2) from the source set of probability space, should *not be empty*, which corresponds $I_{ap} \neq 0$.

Definition1. Information, selected by formula (A0.2) from the source set of probability space, is *not an empty subset of probability space*, which chooses only *a non-repeating (novel) subset* form the source.

Definition 2. Numerical value (in Nat, or Bit), measured by formula (A0.2), determines the *quantity* of information selected from the source.

While, the notion of information formally separates the *distinguished from the undistinguished* subsets (events, processes), associated with the source's related transformation, formula (A0.2) *evaluates numerically this separation.*

From these definitions it follows that both information, as a *subset* (a string) in probability space, and its measure are *mathematical* entities, *resulting* generally from logical operations, while the procedure of both *delivering* a source and the *selection* from the source *does not define* the notion of information.

Some of these operations, as well as transmission and acquisition of information, require spending energy, that leads to binding information with energy and to consideration of related physical, thermodynamic substances, associated with this conversion. A classical physical measurement of this information value (or a "barrier" of the above separation) requires an energy coast to decrease the related physical entropy at least by $k_B$ (at temperature $T$, $k_B$ is Boltzmann constant), which is used for transforming the information measure to the physical one, or for the erasure of this information. A quantum measurement involves the collapse of a wave function needed to transform the information *barrier*, separating the distinguished from the undistinguished subsets, to a quantum physical level, which corresponds to an interaction, or an observation. In both cases, such operations bind the required equivalent energy with information.

The mathematical definition of information (A.02) can be used to define its mathematical equivalent of energy:

$$E_{ap} = I_{ap} k_B. \quad \text{(A.02a)}$$

The information subset, following from the definitions, is standardized by encoding this information in different form of information language, assigning a specific code word from some chosen language's alphabet to each sequence of the subset's symbols. An o*ptimal* encoding of different kind of objects allows transforming their encoded characters in *information* form with preservation the information quantity. The encoding specific features of different observed processes by a common (an optimal) code brings an unique opportunity of revealing the process' connections to its information measure in the related specific code. For example, DNA code encodes amino acids, which establishes the connection between biological nature of the encoded process and its information measure in bites.

The length of the shortest representation of the information string by a *code* or a *program* is the subject of *algorithmic information theory* (AIT). "AIT is the result of putting Shannon's information theory and Turing's computability theory to measure the complexity of an object by the size in bits of the smallest program for computing it" [7].

Encoding (12.3,12.5) provides a *shortest optimal program* (in bits) for each *IPF segment* of *the information process* that connects the considered approach to AIT and Burgin results [6].

**A.1.** Let us show that a control jump, changing the sign of some matrix elements, specifically performing $a_{ii}(v_i) \to -a_{ii}(-v_i), a_{ki}(v_k) \to -a_{ki}(-v_k)$ for a real matrix $A = \begin{pmatrix} a_{ii}, a_{ik} \\ a_{ki}, a_{kk} \end{pmatrix}$, leads to appearance of complex eigenvalues in this matrix.

*Example* 1. Let's have matrix $A = \begin{pmatrix} +3, -2 \\ -4, +1 \end{pmatrix}$ with eigenvalues $\lambda_{1,2} = 2 \pm 3$. After a single dimensional control's changes of the corresponding matrix elements, we get $A = \begin{pmatrix} -3, -2 \\ -4, +1 \end{pmatrix}$, and the matrix eigenvalue become complex $\lambda_{1,2} = 2 \pm j$.

More commonly, considering the dynamic model at the moment $\tau_k^1$ of the equalization of some matrix' eigenvalues according to (3.12), and after reaching a punched locality at $\tau_{k+1} = \tau_k^1 + o$, we get $\lambda_i(\tau_{k+1}) = \lambda_i^*(\tau_{k+1}) = \alpha_i(\tau_{k+1})$; then at the moment $\tau_{k+1}^o$, the control changes the sign of model's matrix, leading to $\lambda_i(\tau_{k+1}^o) = -\lambda_i^*(\tau_{k+1}^o) = \beta_i(\tau_{k+1}^o)$, i.e. the model's eigenvalues become imaginary.

*Remark.* Solving (7.6) at condition $\mathrm{Re}\,\lambda_i\ (\tau_{k-o}) = 0$ brings for an invariant $\mathbf{b}_0^{'} = \beta_{io}\tau_{k-o}^{'}$ the equation

$$2\cos(\gamma \mathbf{b}_0^{'}) - \gamma \sin\gamma(\mathbf{b}_0^{'}) - \exp(\mathbf{b}_0^{'}) = 0,$$

which identifies an interval $t_k^{'}$,when a control, applied at the interval beginning $\tau_k^{'o}$, theoretically, would be able turning to zero the matrix' *real* components by the interval end $\tau_{k-o}^{'}$.

**A.2.** *Example* 2.

Let us have a dynamic model with an imaginary initial eigevalues $\lambda_i(\tau_k^o) = \pm j\beta_{io}$ at a fixed $\beta_{io} \neq 0$.

Following (7.6) we can write the solutions:

$$\lambda_i(\tau_i^1) = j\beta_{io}\exp(j\beta_{io}\tau_k^1)[2 - \exp(-j\beta_{io}\tau_k^1)]^{-1}, \lambda_i^*(\tau_k^1) = -j\beta_{io}\exp(-j\beta_{io}\tau_k^1)[2 - \exp(-j\beta_{io}\tau_k^1)]^{-1}$$

and select the solution's real and imaginary components at the moment $\tau_k^1$:

$$\mathrm{Re}\,\lambda_i(\tau_k^1) = -2\beta_{io}\sin(\beta_{io}\tau_k^1)[5 - 4\cos(\beta_{io}\tau_k^1)]^{-1} = \mathrm{Re}\,\lambda_i^*(\tau_k^1),$$

$$\mathrm{Im}\,\lambda_{it}(\tau) = \pm\beta_{io}(2\cos(\beta_{io}\tau) - 1)[(5 - 4\cos(\beta_{io}\tau)]^{-1}.$$

Applying condition $\mathrm{Im}\,\lambda_i(\tau_k^1) = 0$ (6.10b) to the above solutions, we get $2\cos(\beta_{io}\tau_k^1) - 1 = 0$ and $\beta_{io}\tau_k^1 = \pi / 6 = inv \cong 0.5235$, which corresponds to the specific invariant $\mathbf{b}_o$ in (7.8); at given $\beta_{io}$, $\mathbf{b}_o$ defines $\tau_k^1 = (\beta_{io})^{-1}\pi / 6$. This also allows finding $\mathrm{Re}\,\lambda_i(\tau_k^1) = -0.577\beta_{io}$ and the macrostate at this moment: $x_i(\tau_k^1) = x_i(\tau_k^o)[2 - \exp(-0.577\beta_{io}\tau_k^1)] \cong 1.26\tau_k^1$.

**A.3**. *Example 3*

Let us start with the second order stochastic equation as the microlevel model at $t \in [s, T]$:

$$d\tilde{x}_1(t,\bullet)=a_1(t,\tilde{x}_t,u)dt+\sigma_{11}(t)d\xi_1(t,\bullet)+\sigma_{12}(t)d\xi_2(t,\bullet),\ \tilde{x}_1(s,\bullet)=\tilde{x}_{1s},$$
$$d\tilde{x}_2(t,\bullet)=a_2(t,\tilde{x}_t,u)dt+\sigma_{21}(t)d\xi_1(t,\bullet)+\sigma_{22}(t)d\xi_2(t,\bullet),\tilde{x}_2(s,\bullet)=\tilde{x}_{2s}. \quad \text{(A1)}$$

Suppose the task at the macrolevel is given by a constant vector $\overline{x}_t^1=\overline{x}_o^1 \forall t\in\Delta$, $\overline{x}_o^1\in R^2$, which is chosen to be a beginning of considered coordinate system ($0\overline{x}_1\,\overline{x}_2$).

Then, at $x_t=\overline{x}_t-\overline{x}_t^1$, (sec.1), we have $x_t=\overline{x}_t$ .

The macrolevel model $\dot{x}_t=A_t(x_t+v_t)$ requires the identification of matrix $A_t$ using

$$R_v^1(\tau)=1/2\dot{r}_1(\tau)r_v(\tau)^{-1},\dot{r}_1(\tau)=2b(\tau),\ \tau=(\tau_o,\tau_1),\ \tau_o=s+o, \quad \text{(A2)}$$

which we specify by the following equations:

$$\dot{x}_1(t,\bullet)=A_{11}(t)(x_1(t,\bullet)+v_1(t,\bullet))+A_{12}(t)(x_2(t,\bullet)+v_2(t,\bullet)),\ x_1(s,\bullet)=x_{1s},$$
$$\dot{x}_2(t,\bullet)=A_{21}(t)(x_1(t,\bullet)+v_1(t,\bullet))+A_{22}(t)(x_2(t,\bullet)+v_2(t,\bullet)),x_2(s,\bullet)=x_{2s},$$
$$x_t=\begin{pmatrix}x_1(t,\bullet)\\x_2(t,\bullet)\end{pmatrix},\ v_t=\begin{pmatrix}v_1(t,\bullet)\\v_2(t,\bullet)\end{pmatrix},A(t)=\begin{pmatrix}A_{11},A_{12}\\A_{21},A_{22}\end{pmatrix}, \quad \text{(A3)}$$
$$b=\begin{pmatrix}b_{11},b_{12}\\b_{21},b_{22}\end{pmatrix}=\frac{1}{2}\begin{pmatrix}(\sigma^2{}_{11}+\sigma^2{}_{12}),(\sigma_{11}\sigma_{21}+\sigma_{12}\sigma_{22})\\(\sigma_{11}\sigma_{21}+\sigma_{12}\sigma_{22}),(\sigma^2{}_{21}+\sigma^2{}_{22})\end{pmatrix},$$
$$\dot{r}_1(\tau)=\begin{pmatrix}(E_\tau[\dot{x}_1(x_1(t,\bullet)+v_1(t,\bullet))]),(E_\tau[\dot{x}_1(x_2(t,\bullet)+v_2(t,\bullet))])\\(E_\tau[\dot{x}_2(x_1(t,\bullet)+v_1(t,\bullet))]),(E_\tau[\dot{x}_2(x_2(t,\bullet)+v_2(t,\bullet))])\end{pmatrix}=2\begin{pmatrix}b_{11}(\tau),b_{12}(\tau)\\b_{21}(\tau),b_{22}(\tau)\end{pmatrix},$$
$$r_v(\tau)=\begin{pmatrix}(E_\tau[(x_1(t,\bullet)+v_1(t,\bullet))]^2),(E_\tau[(x_1(t,\bullet)+v_1(t,\bullet))(x_2(t,\bullet)+v_2(t,\bullet))])\\(E_\tau[(x_2(t,\bullet)+v_2(t,\bullet))(x_1(t,\bullet)+v_1(t,\bullet))]),(E_\tau[(x_2(t,\bullet)+v_2(t,\bullet))^2])\end{pmatrix},$$
$$E[\bullet]=\left\{\begin{matrix}E_o=\int_{R^2}[\bullet]P_o(\tau_o,x)dx,x=x_{0,},t\in[\tau_o,\tau_1),\\E_{\tau=\tau_1}=\int_{R^2}[\bullet]P_{\tau_1}(\tau_1,x)dx,x=x_{\tau_1,},t\in[\tau_1,T)\end{matrix}\right\}. \quad \text{(A4)}$$

The process observation is a discrete time process with the elements of $A_t$ as the piece-wise functions of time, which are fixed within each of two discrete intervals $[\tau_o,\tau_1),(\tau_1,T]$, while the controls $v_t(\tau\pm o)=-2x_t(\tau\pm o)$ are applied at the localities of these moments:

$$\tau\pm o=(\tau_o^o=(\tau_o+o),\tau_o^1=(\tau_1-o),\tau_1^1=(\tau_1+o),\tau_1^1=(T-o)). \quad \text{(A5)}$$

This *identification problem* consists of the restoration $A_t(\tau)$ by known $b(\tau)$, $r_1(\tau)$, using here the model solutions $x_t(t,\bullet)$ (with the applied controls $v_t(t,\bullet)$), which are considered to be the equivalents of the process under observation. Using (A4),(A5), we write (A2) via these solutions in matrix forms: $r_v(t)=E[(x_t(t,\bullet)+v_t(t,\bullet))(x_t(t,\bullet)+v_t(t,\bullet))^T]$ and

$$\dot{r}_1(t)=E[\dot{x}_t(t,\bullet)(x_t(t,\bullet)+v_t(t,\bullet))^T+(x_t(t,\bullet)+v_t(t,\bullet))\dot{x}_t(t,\bullet)^T]$$
$$=A_tE[(x_t(t,\bullet)+v_t(t,\bullet))(x_t(t,\bullet)+v_t(t,\bullet))^T]+E[(x_t(t,\bullet)+v_t(t,\bullet))(x_t(t,\bullet)+v_t(t,\bullet))^T]A_t,$$

which at $t=\tau$, $v_t(\tau)=0$, acquire the view

$$r_v(\tau)=E[x_t(\tau)(x_t(\tau)^T]=r(\tau)\ \dot{r}_1(\tau)=A_t(\tau)r(\tau)+r(\tau)A_t(\tau). \quad \text{(A6)}$$

After substitution these relations in (A2), at symmetrical $r(\tau)$, $A_t(\tau)$, we come to

$R_v^1(\tau) = A_t(\tau)$ or $R_v^1(\tau) = b(\tau)r^{-1}(\tau) = A_t(\tau),$ (A7)

which validate the correctness of the identification relation (A2). It's seen that $R_v^1(\tau)$ does not depend on the probability distributions of an *initial* state vector, which is important for many applications.
Let us have the matrix, identified at the first moment $\tau_o$:

$A(\tau_o) = \begin{pmatrix} 2,3 \\ 3,10 \end{pmatrix}$ with eigenvalues $\lambda_1^0$=11, $\lambda_2^0$=1.

At $x_t + v_t = y_t$, the fundamental system and general solutions of (A3**)** within interval $t \in [\tau_o, \tau_1)$ under control $v_t(\tau_o^o, \tau_o^1)$ have the forms:

$Y_t = \begin{pmatrix} y_{11}(t), y_{12}(t) \\ y_{21}(t), y_{22}(t) \end{pmatrix} = \begin{pmatrix} \exp(11t), 3\exp(11t) \\ -3\exp(11t), \exp(t) \end{pmatrix}$, $x_0 = \begin{pmatrix} x_{10} \\ x_{20} \end{pmatrix}$, $y_0 = - x_0$,

$y_1(t,\bullet) = C_1 \exp(11t) - 3C_2 \exp(t)$, $y_2(t,\bullet) = 3C_1 \exp(11t) + C_2 \exp(t)$.
Using the following initial conditions and constants $C_1, C_2$:
$C_1 - 3C_2 = - x_{10}$, $3C_1 + C_2 = - x_{20}$, $C_1 = - 0.1(x_{10} + 3x_{20})$, $C_2 = 0.1(3x_{10} - x_{20})$,
we get the solution of Caushy problem in the form
$y_1(t,\bullet) = - 0.1[x_{10}(\exp(11t) + 9\exp(t)) + 3x_{20}(\exp(11t) - \exp(t))]$,
$y_2(t,\bullet) = - 0.1[3x_{10}(\exp(11t)) - \exp(t)) + x_{20}(9\exp(11t) + \exp(t))]$.
Then we find the moment $\tau_1$ of switching the optimal control using condition (3.11) in the form:

$$\frac{\dot{x}_1(t_1)}{x_1(t_1)} = \frac{11C_1\exp(11t_1) - 3C_2\exp(t_1)}{-v_1(\tau_o,\bullet) + C_1\exp(11t_1) - 3C_2\exp(t_1)} = \frac{\dot{x}_2(t_1)}{x_2(t_1)} = \frac{33C_1\exp(11t_1) + C_2\exp(t_1)}{-v_2(\tau_o,\bullet) + 3C_1\exp(11t_1) + C_2\exp(t_1)}, v_1(\tau_o^o) = -2x_1(\tau_o),$$

$v_2(\tau_o) = -2x_2(\tau_o)$. (A8)

We get equation

$$5\exp(11t_1) - 11\exp(10t_1) + 1 = 0, t_1 = \tau_o^1 - \tau_o^o > 0 \quad \text{(A8a)}$$

having the unique root $\tau_1 \cong$0.7884. These relations illustrate the independency of the discrete moment on a chosen coordinate system. The model solution within interval $t \in (\tau_1, T]$ have the forms

$$x_1(t) = (2 - \exp(0.7t)x_1(\tau_1); x_2(t) = (2 - \exp(0.7)x_2(\tau_1). \quad \text{(A8b)}$$

We also obtain the eigenvalues at moments $\tau_1$ and the final T:

$$\lambda_1^1 = \lambda_2^1 \cong 11, T = \tau_1 + \frac{\ln 2}{\lambda_1^1} \cong 0.851,\ \overline{A}(\tau_1) \cong \begin{pmatrix} 11,0 \\ 0,11 \end{pmatrix}. \quad \text{(A9)}$$

If the identified matrix is negative: $A(\tau_o) = \begin{pmatrix} -2,-3 \\ -3,-10 \end{pmatrix}$, then the moment $\tau_1$ is found by analogy:

$$\frac{11C_1\exp(-11\tau_1)}{-v_1(\tau_o) + C_1\exp(-11\tau_1) - 3C_2\exp(-\tau_1)} = \frac{33C_1\exp(-\tau_1) + C_2\exp(-\tau_1)}{-v_2(\tau_o) + 3C_1\exp(-11\tau_1) + C_2\exp(-\tau_1)}. \quad \text{(A9a)}$$

This equality leads to equation
$5\exp(-11t) - 11\exp(-10t) + 1 = 0, t_1 > 0$ having root $\tau_1 \cong$0.193.

The negative eigenvalues at this moment $\tau_1$:

$$\lambda_1^1 = \lambda_2^1 \cong - 0.7 \quad \text{(A9b)}$$

are changed by applying the needle control, which brings

$\lambda_1^1 = \lambda_2^1 \cong 0.7$, with $T = 0.193 + \dfrac{\ln 2}{0.7} \cong 1.187$.

Applying both identification's and model's equations, we can find the model matrix at the moment $\tau_1$ whose components are determined by relations

$$A_{11}(\tau_1) = \frac{2\exp(12\,\tau_1) - 2.2\exp(11\tau_1) - 1.8\exp(\,\tau_1)}{\exp(12\,\tau_1) - 2\exp(11\tau_1) - 2\exp(\,\tau_1) + 4},$$

$$A_{12}(\tau_1) = A_{21}(\tau_1) = \frac{3(\exp(12\,\tau_1) - 2.2\exp(11\,\tau_1) + 0.2\exp(\,\tau_1))}{\exp(12\,\tau_1) - 2\exp(11\,\tau_1) - 2\exp(\,\tau_1) + 4},$$

$$A_{22}(\tau_1) = \frac{10\exp(12\,\tau_1) - 19.8\exp(11\,\tau_1) - 0.2\exp(\,\tau_1)}{\exp(12\tau_1) - 2\exp(11\tau_1) - 2\exp(\,\tau_1) + 4}. \qquad \text{(A10)}$$

The numerical solutions for this matrix are:

$$A\,(\tau_1) \cong \begin{pmatrix} 11.006, -0.00077 \\ -0.00077, 11.004 \end{pmatrix} \cong \begin{pmatrix} 11, 0 \\ 0, 11 \end{pmatrix}. \qquad \text{(A10a)}$$

Comparing both results for $A\,(\tau_1)$ (A9) and (A10a) we come to conclusion that we have identified $A_t \forall t \in (\,\tau_1, T)$ with a high precision (defined by a computation accuracy), which does not depend on a chosen coordinate system.
The step-wise control provides the changing of matrix $A(\tau)$ sign at any of $\tau$ -localities.

The optimal processes within the discrete interval $t \in (\tau_1, T)$ with matrix' eigenvalues (A9c):

$$x_1(t) = (2 - \exp(11t)x_1(\tau_1); x_2(t) = (2 - \exp(11t)x_2(\tau_1),$$

are distinctive only by the starting states ( $x_1(\tau_1)$, $x_2(\tau_1)$ ). Analogous form has optimal processes with the matrix negative eigenvalues in (A9):

$$x_1(t) = (2 - \exp(-0.7t)x_1(\tau_1); x_2(t) = (2 - \exp(-0.7t)x_2(\tau_1).$$

Therefore, the matrix' identification proceeds *during the optimal control action* at each extremal segment.
Let us determine the phase trajectories of the dynamic model at both discrete intervals.
At first, we will find these trajectories for the diagonalized system at $t \in [\tau_o, \tau_1)$:

$\dfrac{dz_1}{dz_2} = -\dfrac{-\lambda^o{}_1}{-\lambda^0{}_2}\dfrac{z_1}{z_2}$, $z_2 = \pm|\iota|\, z_1^{\lambda_2^o/\lambda_1^o}$, $\iota \in R^1$, $\pm|\iota| = z_2(\tau_o) / z_1(\tau_o)^{\lambda_2^o/\lambda_1^o}$. The phase trajectories of this system (Fig.A.1a) present the $\iota$ -parametrical family of the curves with a singular tangle-point in (0,0) (a "knot").

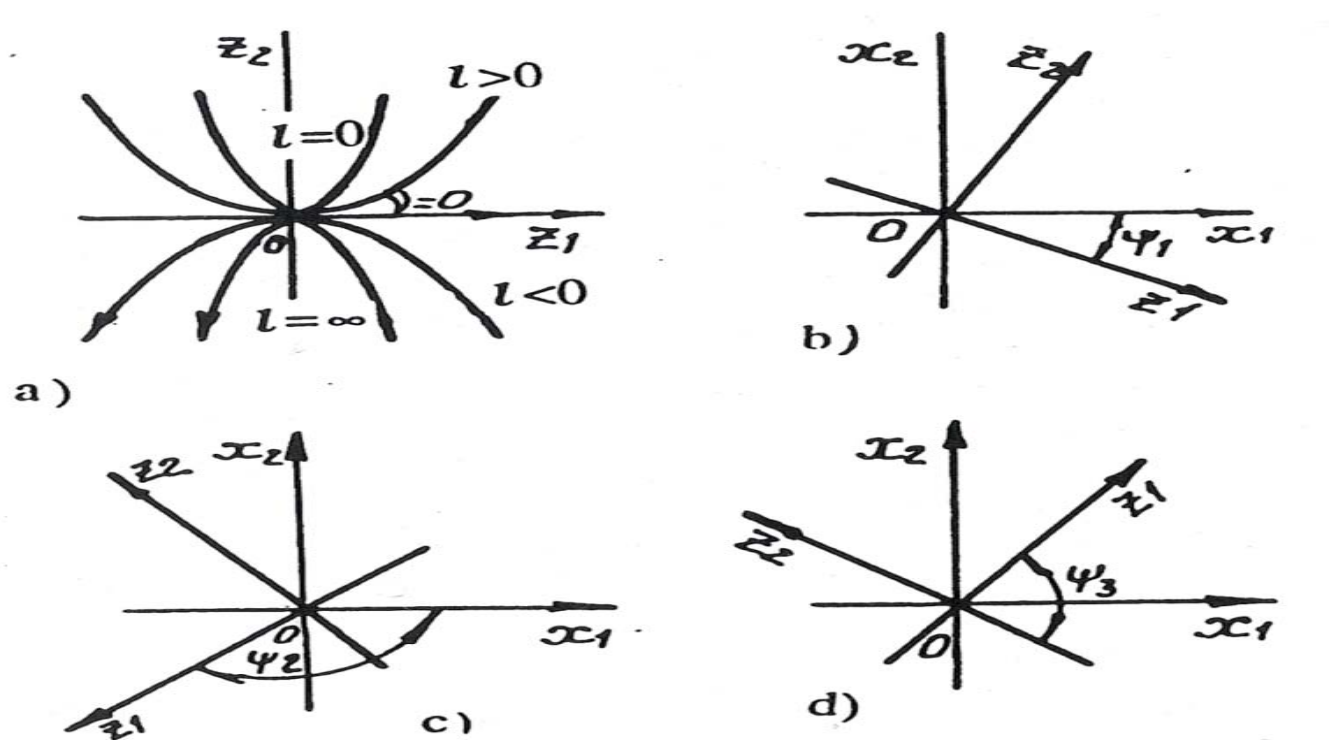

Fig.A.1(a-c). The phase pictures of dynamic model in the initial coordinate system at the first discrete interval.

The model phase picture in the initial coordinate system ($0x_1\,x_2$) (Fig.A.1b-d) we get after turning the coordinate system on Figs.A.1a (with the phase picture) on the angles, determined by the matrix's identified components (A10). To get the phase trajectories at the second discrete interval, we will find the phase picture of relation (A9a).
We come to

$$a_{11}(x_1+v_1)x_2+a_{12}(x_2+v_2)x_2=a_{21}(x_1+v_1)x_1+a_{22}(x_2+v_2)x_1;\,a_{12}=a_{21}; \qquad \text{(A11)}$$

$$a'_{11}x_1^2+2a'_{12}x_1x_2+a'_{22}x_2^2+2a'_{13}x_1+2a'_{23}x_2+a'_{23}=0,$$

where $a'_{11}=a_{21};\,a'_{12}=1/2(a_{22}-a_{11});\,a'_{22}=-a_{12};\,a'_{13}=1/2(a_{21}v_1-a_{22}v_2);$

$a'_{23}=-1/2(a_{11}v_1+a_{12}v_2),\,a'_{33}=0$ .

The equation of a second order for a line (A11) we will transform to a canonic form using equation

$$a'_{11}(x'_1)^2+2a'_{12}x'_1x'_2+a'_{22}(x'_2)^2+\frac{I_3}{I_2}=0, \qquad \text{(A11a)}$$

where $x=x'_1+x^o$ and $x^o=(x_1^o,x_2^o)$ are the coordinates of the beginning of system coordinates ($0x'_1\,x'_2$) being transformed into the system coordinates ($0x_1\,x_2$), which satisfy the equations $a'_{11}x_1^o+a'_{12}x_2^o+a'_{13}=0,\,a'_{12}x_1^o+a'_{22}x_2^o+a'_{23}=0$, with the parameters of transformation

$$I_2=det\begin{pmatrix}a'_{11},a'_{12}\\ a'_{12},a'_{22}\end{pmatrix}=det\begin{pmatrix}a_{12},1/2(a_{22}-a_{11})\\ 1/2(a_{22}-a_{11}),-a_{12}\end{pmatrix}=inv,\ I_2=-\,25,$$

$$I_3=det\begin{pmatrix}a'_{11},a'_{12},a'_{13}\\ a'_{12},a'_{22},a'_{23}\\ a'_{13},a'_{23},a'_{33}\end{pmatrix}=det\begin{pmatrix}3,4,1/2(3v_1+10v_2)\\ 4,-3,-1/2(3v_1+2v_2)\\ 1/2(3v_1+10v_2),-1/2(3v_1+2v_2),0\end{pmatrix},$$

$$I_3=-1/4\,(33v^2{}_1+327v^2{}_2+196v_1\,v_2)=inv,\ \frac{I_3}{I_2}=-\,0.01(33v^2{}_1+327v^2{}_2+196v_1\,v_2). \qquad \text{(A12)}$$

After a simplification we obtain the equations

$$a''_{11}(x''_1)^2+a''_{22}(x''_2)^2+\frac{I_3}{I_2}=0,\ \begin{pmatrix}x''_1\\ x''_2\end{pmatrix}=\begin{pmatrix}\cos\vartheta,\sin\vartheta\\ -\sin\vartheta,\cos\vartheta\end{pmatrix}\begin{pmatrix}x'_1\\ x'_2\end{pmatrix},\ ctg2\,\vartheta=\frac{a'_{11}-a'_{22}}{2a'_{12}}=0.75;$$

$$I_2=inv<0;\ I_2=det\begin{pmatrix}a''_{11},a''_{12}\\ a''_{12},a''_{22}\end{pmatrix}=det\begin{pmatrix}a''_{11},0\\ 0,a''_{22}\end{pmatrix}=a''_{11}a''_{22}<0\,,$$

$$a''_{11}=a'_{12}\sin 2\,\vartheta+\frac{a'_{11}-a'_{22}}{2}\cos 2\,\vartheta\,,\ a''_{11}\overset{def}{>}0,\ a''_{22}=-a'_{12}\sin 2\,\vartheta-\frac{a'_{11}-a'_{22}}{2}\cos 2\,\vartheta\,,\ a''_{22}\overset{def}{<}0..$$

At $I_3$<0 we get the canonical equation of a hyperbola with respect to the real axis ($0'\,x''_1$) and the imaginary axis– ($0'\,x''_2$) (Fig. A.2a):

$$\frac{(x''_1)^2}{[(\frac{I_3}{I_2a''_{11}})^{\frac{1}{2}}]^2}-\frac{(x''_2)^2}{[(\frac{I_3}{-a''_{22}I_2})^{\frac{1}{2}}]^2}=1. \qquad \text{(A13)}$$

At $I_3$>0 we come to other hyperbola on Fig.A.2a, with respect to the real axis ($0'\,x''_2$) and the imaginary axis– ($0'\,x''_1$):

$$\frac{(x_1^{''})^2}{[(-\frac{I_3}{I_2 a^{''}_{11}})^{\frac{1}{2}}]^2} - \frac{(x_2^{''})^2}{[(\frac{-I_3}{-a^{''}_{22} I_2})^{\frac{1}{2}}]^2} = -1. \quad \text{(A13a)}$$

At $I_3$=0 we get a couple of the equations that satisfy to the coordinates of the points, located on the straight lines, represented the asymptotes of the hyperbolas (A13),(A13a) (Figs. A.2b,c):

$$\frac{x_1^{''}}{(\frac{1}{a^{''}_{11}})^{\frac{1}{2}}} + \frac{x_2^{''}}{(\frac{1}{-a^{''}_{22}})^{\frac{1}{2}}} = 0, \quad \frac{x_1^{''}}{(\frac{1}{a^{''}_{11}})^{\frac{1}{2}}} - \frac{x_2^{''}}{(\frac{1}{-a^{''}_{22}})^{\frac{1}{2}}} = 0. \quad \text{(A14)}$$

On the coordinate plane $(0^{'}\, x_1^{''}\, x_2^{''})$, the phase picture of relation (A8) represents a couple of the conjugated hyperbolas with the asymptotes, defined by equation (A14) and a saddle singular point (0,0) (Fig. A.2d).
The phase trajectories of the dynamic system at the second discrete interval, after switching the control, are

$$\dot{y}_1(t,\bullet) = A_{11}(\tau_1+0)y_1(t,\bullet) = \lambda_1^1 y_1(t,\bullet),\ t\in(\tau_1,T),$$

$$\dot{y}_2(t,\bullet) = A_{22}(\tau_1+0)\,y_2(t,\bullet) = \lambda_2^1\, y_2(t,\bullet),\ \lambda_1^1 = \lambda_2^1;\ \frac{dy_1}{dy_2} = \frac{y_1}{y_2}, y_1 = \pm|C|y_2,\ C\in R^1,$$

$$\pm|C| = \frac{y_1(\tau_1,\bullet)}{y_2(\tau_1,\bullet)} = \frac{x_1(\tau_1,\bullet)}{x_2(\tau_1,\bullet)};\ x_1 + v_1 = \frac{x_1(\tau_1,\bullet)}{x_2(\tau_1,\bullet)}(x_2+v_2).$$

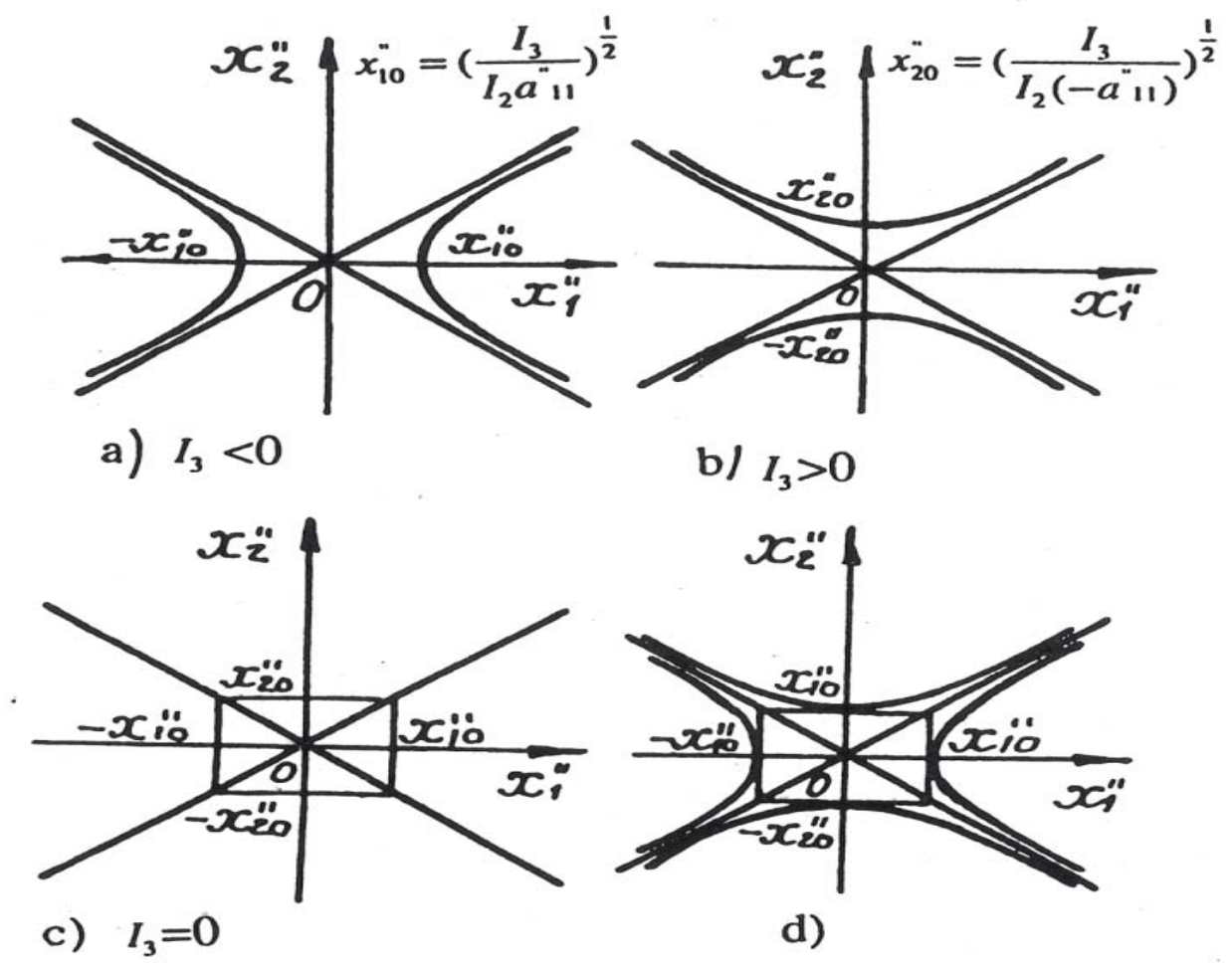

Fig.A.2. The phase pictures of the dynamic model (a-c) and the relation (A8) (d) at the second discrete interval after the transforming to the initial coordinate system.

The phase picture at $t\in(\tau_1,T)$ presents a family of the straight lines

$$x_2 = \frac{\dot{x}_1(\tau_1,\bullet)}{x_1(\tau_1,\bullet)}x_1 \quad \text{(A15)}$$

with parameter $\frac{x_1(\tau_1,\bullet)}{x_2(\tau_1,\bullet)}$. The phase picture of equality $\frac{\dot{x}_1(\tau_1,\bullet)}{x_1(\tau_1,\bullet)} = \frac{\dot{x}_2(\tau_1,\bullet)}{x_2(\tau_1,\bullet)}$, $t\in(\tau_1,T)$ has the form

$$A_{22}(\tau_1+0)\,v_2\,(\tau_1+0)\,x_1\,(t,\bullet) = A_{11}(\tau_1+0)\,v_1\,(\tau_1+0)\,x_2\,(t,\bullet),\ x_1 = \frac{x_1(\tau_1,\bullet)}{x_2(\tau_1,\bullet)}\,x_2. \quad \text{(A16)}$$

At the second discrete interval, the phase pictures of the dynamic model (A16) and relation (A15) coincide.
The comparison of the Figs.A.1,A.2 for (A15),(A16) illustrates the geometrical interpretation of the constraint action.

At the moment of applying the control, the phase pictures of the dynamic model and relation (A15) are changed by the jumps. This leads to the *renovation* of matrix $A(\tau_1+0)$ with respect to matrix $A(\tau_1-0)$, and it creates the new model's peculiarities. Let us find the jump of the phase speed at $\tau_1$:

$$\delta\dot{x}(\tau_1,\bullet)=\dot{x}(\tau_1+0,\bullet)-\dot{x}(\tau_1-0,\bullet)=-A(\tau_1+0)x(\tau_1,\bullet)-A(\tau_1-0)(x(\tau_1,\bullet)+v(\tau_o^o))$$

$$=A(\tau_1+0)+A(\tau_1-0))x(\tau_1,\bullet)+2A(\tau_1-0)x_0 .$$

From the phase speed's expressions and the previous relations we have

$$A(\tau_1+0)+A(\tau_1-0)\cong\begin{pmatrix}13,3\\3,21\end{pmatrix},\ \delta\dot{x}(\tau_1,\bullet)=Kx_0,$$

$K_{11}$=2.2exp(11 $\tau_1$)+10.8exp( $\tau_1$), $K_{12}=K_{21}$=6.6exp(11 $\tau_1$)− 3.6exp( $\tau_1$),
$K_{22}$=19.8exp(11 $\tau_1$)+1.2exp( $\tau_1$)− 22.

At $\tau_1$=0.7884, we obtain the numerical results: $K=\begin{pmatrix}12848.65,38532.75\\38532.75,115602.66\end{pmatrix}$, which determine the values of both jumps:

$\delta\dot{x}_1(\tau_1,\bullet)$=51381.4, $\delta\dot{x}_2(\tau_1,\bullet)$=154135.41 at $x_0=\begin{pmatrix}1\\1\end{pmatrix}$.

*Therefore, the changes of model's original nonlinear operator are also identified at the DP* $(\tau_i)$ *by the jump-wise sequence of* $\Delta A(\tau_i)$.

The identification method depends on the accuracies of computing the correlations functions $r(\tau)$ and on the feedback effect in a close-loop control. The identification of the *actual* object's operator under the concurrent close-loop control has been implemented in the *electro-technological process* using a direct measuring of the diffusion conductivity according to [23,26]. It expedites the close-loop dynamics and minimizes the related error, compared to statistical method of computing the correlation function. This error was not exceeded 10%.

For the solution of the *consolidation* problem, we rotate initial coordinate system on angle $\varphi$ to find such a coordinate system (0 $z_1'$ $z_2'$), where the optimal processes are undistinguished. Using the relations for consolidation, we get

$$\varphi_{12}=\varphi=\text{arctg}\left(\frac{x_2(\tau_1,\bullet)-x_1(\tau_1,\bullet)}{x_2(\tau_1,\bullet)+x_1(\tau_1,\bullet)}\right)+k\pi,\ k=0,\pm1,\pm2,...,$$

$$x(\tau_1,\bullet)=\begin{pmatrix}x_1(\tau_1,\bullet)\\x_2(\tau_1,\bullet)\end{pmatrix}=\begin{pmatrix}5839.294,17511.88\\17511.88,52537.66\end{pmatrix}\begin{pmatrix}1\\1\end{pmatrix}=\begin{pmatrix}23351.17\\70049.54\end{pmatrix},\ \text{at } x_0=\begin{pmatrix}1\\1\end{pmatrix}$$

and the angle $\varphi=\text{arctg}\dfrac{46698.37}{93400.71}+k\pi\cong\text{arctg}0.5+k\pi$, $\varphi|_{k=0}\cong0.1472\pi$. This allows us to characterize each pair $x_1(\tau_1)=x_2(\tau_1)$ of the equal state vectors by a *single joint* vector. The consequent realization of this process for a multi-dimensional system leads to *building* the object's cooperative information network (IN), considered in [23,26].